\documentstyle[12pt]{article}
\begin{document}

\def\ad{^\dagger }
\def\half{{1\over 2}}
\def\ho{\hbar\omega }
\def\lg{{\langle}}
\def\mt{\mapsto }
\def\od{\odot }
\def\ot{\otimes }
\def\ra{{\rightarrow}}
\def\rg{{\rangle}}
\def\st{{\sqrt 2}}
\def\ta{\theta }
\def\v{{\,|\,}}
\def\Re{\hbox{Re}}
\def\Tr{\hbox{Tr}}

\def\A{{\cal A}}
\def\B{{\cal B}}
\def\BT{{\tilde{\cal B}}}
\def\C{{\cal C}}
\def\D{{\cal D}}
\def\E{{\cal E}}
\def\F{{\cal F}}
\def\G{{\cal G}}
\def\H{{\cal H}}
\def\HB{{\breve{\cal H}}}
\def\IB{\breve{I}}
\def\J{{\cal J}}
\def\P{{\cal P}}
\def\Q{{\cal Q}}
\def\SS{{\cal S}}
\def\W{{\cal W}}
\def\X{{\cal X}}
\def\Z{{\cal Z}}

\def\At{{\tilde A}}
\def\Ct{{\tilde C}}
\def\Dt{{\tilde D}}
\def\Ft{{\tilde F}}
\def\Mt{{\tilde M}}
\def\Pt{{\tilde P}}
\def\Qt{{\tilde Q}}
\def\Rt{{\tilde R}}

\title{Consistent Histories and  Quantum Reasoning}

\author{Robert B. Griffiths\thanks{Electronic mail: rgrif@cmu.edu}\\
Department of Physics\\ Carnegie Mellon University\\ Pittsburgh, PA 15213,
U.S.A.}

\date{Version of 4 June 1996}
\maketitle

\begin{abstract}
	A system of quantum reasoning for a closed system is developed by
treating non-relativistic quantum mechanics as a stochastic theory.  The sample
space corresponds to a decomposition, as a sum of orthogonal projectors, of the
identity operator on a Hilbert space of histories.  Provided a consistency
condition is satisfied, the corresponding Boolean algebra of histories, called
a {\it framework}, can be assigned probabilities in the usual way, and within a
single framework quantum reasoning is identical to ordinary probabilistic
reasoning.  A refinement rule, which allows a probability distribution to be
extended from one framework to a larger (refined) framework, incorporates the
dynamical laws of quantum theory.  Two or more frameworks which are
incompatible because they possess no common refinement cannot be simultaneously
employed to describe a single physical system.
	Logical reasoning is a special case of probabilistic reasoning in which
(conditional) probabilities are 1 (true) or 0 (false).  As probabilities are
only meaningful relative to some framework, the same is true of the truth or
falsity of a quantum description.
	The formalism is illustrated using simple examples, and the physical
considerations which determine the choice of a framework are discussed.
\end{abstract}

			\section{Introduction}
\label{intro}

	Despite its success as a physical theory, non-relativistic quantum
mechanics is beset with a large number of conceptual difficulties.  While the
mathematical formalism is not at issue, the physical interpretation of this
formalism remains controversial.  Does a wave function describe a physical
property of a quantum system, or is it merely a means for calculating
something?  Do quantum measurements reveal pre-existing properties of a
measured system, or do they in some sense create the properties they reveal?
These are but two of the questions which trouble both beginners and experts.

	It would be wrong to dismiss these issues as mere ``philosophical
problems''.  The effective use of a mathematical structure as part of a
physical theory requires an intuitive understanding of what the mathematics
means, both in order to relate it to the real world of laboratory experiment,
and in order to motivate the approximations which must be made when the exact
solution of some equation is a practical impossibility.  In older domains of
application of quantum theory, such as scattering theory, there is by now a
well-developed set of rules, and while the justification for these is somewhat
obscure, once they have been learned, they can be applied without worrying too
much about ``what is really going on''.  But when quantum mechanics is applied
in an unfamiliar setting, such as is happening at the present time in the field
of quantum computation \cite{dv95}, its unresolved conceptual difficulties are
a serious impediment to physical understanding, and advances which enable one
to think more clearly about the problem can lead to significant improvements in
algorithms, as illustrated in
\cite{gn96}.

	The principal thesis of the present paper is that the major conceptual
difficulties of non-relativistic quantum theory (which, by the way, are also
present in relativistic theories) can be eliminated, or at least tamed, by
taking a point of view in which quantum theory is fundamentally a {\it
stochastic} theory, in terms of its description of the time development of a
physical system.  The approach found in typical textbooks is that the time
development of a quantum system is governed by a deterministic Schr\"odinger
equation up to the point at which a measurement is made, the results of which
can then be interpreted in a probabilistic fashion.  By contrast, the point of
view adopted here is that a quantum system's time evolution is fundamentally
stochastic, with probabilities which can be calculated by solving
Schr\"odinger's equation, and deterministic evolution arises only in the
special case in which the relevant probability is one.  This approach makes it
possible to recover all the results of standard textbook quantum theory, and
much else besides, in a manner which is conceptually much cleaner and does not
have to make excuses of the ``for all practical purposes'' variety, justly
criticized by Bell \cite{bl90}.

	Most of the tools needed to formulate time development in quantum
theory as a stochastic process have already appeared in the published
literature.  They include the idea that the properties of a quantum system are
associated with subspaces of an appropriate Hilbert space \cite{vn55}, the
concept of a quantum history as a set of events at a sequence of successive
times \cite{gr84}, the use of projectors on a tensor product of copies of the
Hilbert space to represent these histories \cite{is94}, the notion that a
collection of such histories can, under suitable conditions (``consistency''),
form an event space to which quantum theory ascribes probabilities
\cite{gr84,gr93,gr94,om88,om92,om94,gmh90,gmh93}, and rules which restrict
quantum reasoning processes to single consistent families of histories
\cite{om88,om92,om94}.

	The present paper thus represents an extension of the ``consistent
histories'' procedure for quantum interpretation.  The new element added to
previous work is the systematic development of the concept of a {\it
framework}, the quantum counterpart of the space of events in ordinary
(``classical'') probability theory, and the use of frameworks in order to
codify and clarify the process of reasoning needed to discuss the time
development of a quantum system.  A framework is a Boolean algebra of commuting
projectors (orthogonal projection operators) on the Hilbert space of quantum
histories, Sec.~\ref{proj}, which satisfies certain {\it consistency
conditions}, Sec.~\ref{wc}.  Reasoning about how a quantum system develops in
time, Sec.~\ref{qr}, then amounts to the application of the usual rules of
probability theory to probabilities defined on a framework, together with an
additional {\it refinement rule} which permits one to extend a given
probability distribution to a refinement or enlargement of the original
framework, Sec.~\ref{prob}.  In particular, the standard (Born) rule for
transition probabilities in a quantum system is a consequence of the refinement
rule for probabilities.  Logical rules of inference, in this context, are
limiting cases of probabilistic rules in which (conditional) probabilities are
one (true) or zero (false).  Because probabilities can only be defined relative
to a framework, the notions of ``true'' and ``false'' as part of a quantum
description are necessarily framework dependent, as suggested in \cite{gr93b};
this rectifies a problem \cite{dk956} with Omn\`es' approach
\cite{om92,om94} to defining ``truth'' in the context of consistent histories,
and responds to certain objections raised by d'Espagnat
\cite{de87,de89,de90,de95}.

	The resulting structure is applied to various simple examples in
Sec.~\ref{exams} to show how it works.  These examples illustrate how the
intuitive significance of a projector can depend upon the framework in which it
is embedded, how certain problems of measurement theory are effectively dealt
with by a consistent stochastic approach, and how the system of quantum
reasoning presented here can help untangle quantum paradoxes.  In particular, a
recent criticism of the consistent histories formalism by Kent \cite{kt96a},
involving the inference with probability one from the same initial data, but in
two incompatible frameworks, of two events represented by mutually orthogonal
projection operators, is considered in Sec.~\ref{vaidman} with reference to a
paradox introduced by Aharonov and Vaidman \cite{av91}.  For reasons explained
there and in Sec.~\ref{harmony}, such inferences do not, for the approach
discussed in this paper, give rise to a contradiction.

	Since the major conceptual difficulties of quantum theory are
associated with the existence of {\it incompatible} frameworks with no exact
classical analog, Sec.~\ref{interp} is devoted to a discussion of their
significance, along with some comments on how the world of classical physics
can be seen to emerge from a fundamental quantum theory. Finally,
Sec.~\ref{summ} contains a brief summary of the conclusions of the paper,
together with a list of open questions.


			\section{Projectors and Histories}
\label{proj}

	Ordinary probability theory \cite{fl68} employs a {\it sample space}
which is, in the discrete case, a collection of {\it sample points}, regarded
as mutually exclusive outcomes of a hypothetical experiment.  To each sample
point is assigned a non-negative {\it probability}, with the sum of the
probabilities equal to one.  An {\it event} is then a set of one or more sample
points, and its probability is the sum of the probabilities of the sample
points which it contains.  The events, under the operations of intersection and
union, form a {\it Boolean algebra of events}.  In this and the following two
sections we introduce quantum counterparts for each of these quantities.
Whereas in many physical applications of probability theory only a single
sample space is involved, and hence its identity is never in doubt and its
basic properties do not need to be emphasized, in the quantum case one
typically has to deal with many different sample spaces and their corresponding
event algebras, and clear thinking depends upon keeping track of which one is
being employed in a particular argument.

	The quantum counterpart of a sample space is a {\it decomposition of
the identity\/} on an appropriate Hilbert space.  We shall always assume that
the Hilbert space is {\it finite dimensional\/}; for comments on this, see
Sec.~\ref{open}.  On a finite-dimensional space, such a decomposition of the
identity $I$ corresponds to a (finite) collection of orthogonal projection
operators, or {\it projectors} $\{B_i\}$, which satisfy:
\begin{equation}
 I=\sum_i B_i,\quad B_i\ad = B_i,\quad B_i B_j = \delta_{ij} B_i.
\label{e2.1}
\end{equation}
The Boolean algebra $\B$ which corresponds to the event algebra is then the
collection of all projectors of the form
\begin{equation}
 P=\sum_i \pi_i B_i,
\label{e2.2}
\end{equation}
where $\pi_i$ is either 0 or 1; different choices give rise to the $2^n$
projectors which make up $\B$ in the case in which the sum in (\ref{e2.1})
contains $n$ terms.  We shall refer to the $\{B_i\}$ as the {\it minimal
elements} of $\B$.

	For a quantum system at a single time, $I$ is the identity operator on
the usual Hilbert space $\H$ used to describe the system, and projectors of the
form $P$, or the subspace of $\H$ onto which they project, represent {\it
properties} of the system.  (See Sec.~\ref{exams} for some examples.)  	The
phase space of classical Hamiltonian mechanics provides a useful analogy in
this connection.  A {\it coarse graining} of the phase space in which it is
divided up into a number of non-overlapping cells corresponds to (\ref{e2.1}),
where $B_i$ is the {\it characteristic function} of the $i$'th cell, that is,
the function which is $1$ for points of the phase space inside the cell, and
$0$ for points outside the cell, and $I$ the function which is $1$ everywhere.
The events in the associated algebra correspond to regions which are unions of
some collection of cells, and their characteristic functions $P$ again have the
form (\ref{e2.2}).

	Projectors of the form (\ref{e2.2}) corresponding to a particular
decomposition of the identity (\ref{e2.1}) commute with each other and form a
Boolean algebra $\B$, in which the negation of a property, ``not $P$'',
corresponds to the complement
\begin{equation}
 \Pt = I-P
\label{e2.4}
\end{equation}
of the projector $P$, and the meet and join operations are defined by:
\begin{equation}
 P\land Q = PQ,\quad P\lor Q = P + Q -PQ.
\label{e2.5}
\end{equation}
Note that $P\land Q$ corresponds to the conjunction of the two properties:
``$P$ {\it and\/} $Q$'', whereas $P\lor Q$ is the disjunction, ``$P$ {\it or\/}
$Q$''.  Precisely the same definitions (\ref{e2.4}) and (\ref{e2.5}) apply in
the case of characteristic functions for the coarse graining of a classical
phase space, and the intuitive significance is much the same as in the quantum
case.  Of course, two quantum projectors $P$ and $Q$ need not commute with each
other, in which case they cannot belong to the same Boolean algebra $\B$, and
the properties ``$P$ {\it and\/} $Q$'' and ``$P$ {\it or\/} $Q$'' are not
defined, that is, they are meaningless.  (Note that at this point our treatment
diverges from traditional quantum logic as based upon the ideas of Birkhoff and
von Neumann \cite{bvn36}.)

	A {\it history} of a quantum mechanical system can be thought of as a
sequence of properties or {\it events}, represented by projectors $E_1, E_2,
\ldots E_n$ on the Hilbert space $\H$ at a succession of times $t_1 < t_2 <
\cdots t_n$.  The projectors corresponding to different times are not required
to belong to the same Boolean algebra, and need not commute with each other.
Following a suggestion by Isham \cite{is94}, we shall represent such a history
as a projector
\begin{equation}
 Y=E_1\od E_2 \od \cdots \od E_n
\label{e2.6}
\end{equation}
on the {\it history space}
\begin{equation}
 \HB = \H\od \H\od \cdots \od \H
\label{e2.7}
\end{equation}
consisting of the tensor product of $n$ copies of $\H$.  (We use $\od$ in place
of the conventional $\ot$ to avoid confusion in the case in which $\H$ itself
is the tensor product of two or more spaces.) The number $n$ of times entering
the history can be arbitrarily large, but will always be assumed to be finite,
which ensures that $\HB$ is finite-dimensional as long as $\H$ itself is
finite-dimensional.

	The intuitive interpretation of a history of the form (\ref{e2.6}) is
that event $E_1$ occurs in the closed quantum system at time $t_1$, $E_2$
occurs at time $t_2$, and so forth.  The consistent history approach allows a
realistic interpretation of such a history so long as appropriate consistency
conditions, Sec.~\ref{wc}, are satisfied.  Following \cite{is94}, we shall
allow as a possible history {\it any} projector on the space (\ref{e2.7}), and
not only those of the product form (\ref{e2.6}).  The intuitive significance of
such ``generalized histories'' is not clear, because most physical applications
which have appeared in the literature up to the present time employ ``product
histories'' of the form (\ref{e2.6}).

	One sometimes needs to compare two histories $Y_1$ and $Y_2$ defined on
two different sets of times, say $t'_1 < t'_2 <\ldots t'_p,$ and $t''_1 <t''_2
<\ldots t''_q$.  It is then convenient to {\it extend} both $Y_1$ and $Y_2$ to
the collection of times $t_1<t_2<\ldots t_n$ which is the union of these two
sets, by introducing in the product (\ref{e2.6}) the identity operator $I$ on
$\H$ at every time at which the history was not originally defined.  We shall
use the same symbols, $Y_1$ and $Y_2$, for the extensions as for the original
histories, as this causes no confusion, and the physical significance of the
original history and its extension is the same, because the property $I$ is
always true.

	A useful classical analogy of a quantum history is obtained by
imagining a coarse graining of the phase space, and then thinking of the
sequence of cells occupied by the phase point corresponding to a particular
initial state, for a sequence of different times.  One must allow for different
coarse grainings at different times in order to have an analog of the full
flexibility possible in the quantum description.

	A probabilistic description of a closed quantum system as a function of
time can be based upon a Boolean algebra $\F$ of histories generated by a
decomposition of the identity operator $\IB$ on $\HB$:
\begin{equation}
 \IB= \sum_i F_i, \quad F_i\ad = F_i,\quad F_i F_j = \delta_{ij} F_i,
\label{e2.8}
\end{equation}
where the projectors $\{F_i\}$ will be referred to as the {\it minimal
elements} of $\F$.  The different projectors in $\F$ are of the form
\begin{equation}
 Y=\sum_i \upsilon_i F_i,
\label{e2.9}
\end{equation}
with each $\upsilon_i$ either 0 or 1, and the corresponding Boolean algebra is
constructed using the obvious analogs of (\ref{e2.4}) and (\ref{e2.5}).  We
shall refer to $\F$ as a {\it family of histories}, and, when certain
additional (consistency) conditions are satisfied, as a {\it framework}.

		\section{Weights and Consistency}
\label{wc}

	Quantum dynamics is described by a collection of time evolution
operators $T(t',t)$, thought of as carrying the system from time $t$ to time
$t'$, so that a state $|\psi(t)\rg$ evolving by Schr\"odinger's equation
satisfies
\begin{equation}
 |\psi(t)\rg = T(t,0)|\psi(0)\rg.
\label{e3.1}
\end{equation}
We assume that these operators satisfy the conditions:
\begin{equation} 
 T(t,t) = I,\quad T(t'',t')T(t',t)= T(t'',t),\quad T(t',t)\ad = T(t,t'),
\label{e3.2} 
\end{equation}
which, among other things, imply that $T(t',t)$ is unitary.  If the system has
a time-independent Hamiltonian, $T$ takes the form
\begin{equation} 
 T(t',t) = \exp[-i(t'-t) H/\hbar].
\label{e3.3} 
\end{equation}
However, none of the results in this paper depends upon assuming the form
(\ref{e3.3}).

	Given the time transformation operators, we define the {\it weight
operator}
\begin{equation} 
 K(Y) = E_1 T(t_1,t_2) E_2 T(t_2,t_3) \ldots T(t_{n-1},t_n) E_n
\label{e3.4} 
\end{equation}
for the history $Y$ in (\ref{e2.6}).  It is sometimes convenient to define the
{\it Heisenberg projector}
\begin{equation} 
 \hat E_j = T(t_r,t_j) E_j T(t_j,t_r)
\label{e3.5} 
\end{equation}
corresponding to the event $E_j$ at time $t_j$, where $t_r$ is some arbitrary
reference time independent of $j$, and the corresponding {\it Heisenberg weight
operator}
\begin{equation} 
 \hat K(Y) = \hat E_1 \hat E_2 \cdots \hat E_n.
\label{e3.6} 
\end{equation}
For histories which are not of the form (\ref{e2.6}), but are represented by
more general projectors on $\HB$, one can follow the procedure in \cite{is94}
and define a weight operator by noting that (\ref{e3.4}) also makes sense when
the $E_j$ are arbitrary operators (not just projectors), and then use
linearity,
\begin{equation} 
 K(Y' + Y'' + Y''' +\cdots) = K(Y') + K(Y'') + K(Y''') +\cdots,
\label{e3.7} 
\end{equation}
to extend $K$ to a linear mapping from operators on $\HB$ to operators on $\H$.

	Next, we define an inner product on the linear space of operators on
$\H$ by means of
\begin{equation} 
 \lg A,B\rg = \Tr[A\ad B] = \lg B,A\rg^*.
\label{e3.8} 
\end{equation}
In particular, $\lg A,A\rg$ is positive, and vanishes only if $A=0$. In terms
of this inner product we define the {\it weight} of a history $Y$ as
\begin{equation} 
 W(Y) = \lg K(Y),K(Y)\rg = \lg \hat K(Y),\hat K(Y)\rg.
\label{e3.9} 
\end{equation} 
Intuitively speaking, the weight is like an unnormalized probability. If
$W(Y)=0$, this means the history $Y$ violates the dynamical laws of quantum
theory, and thus the probability that it will occur is zero. Next, define a
function
\begin{equation} 
 \theta(X|Y) = W(XY)/W(Y)
\label{e3.10} 
\end{equation}
on pairs of histories $X$ and $Y$, as long as the right side of (\ref{e3.10})
makes sense, that is, $XY=YX$ is a projector, and $W(Y) > 0$.  Under
appropriate circumstances, described in Secs.~\ref{prob} and \ref{qr},
$\theta(X| Y)$, which is obviously non negative, functions as a conditional
probability of $X$ given $Y$, which is why we write its arguments separated by
a bar.

	Let $Y$ and $Y'$ be projectors in the Boolean algebra $\F$ or histories
based upon (\ref{e2.8}). In the analogous classical situation, where $W(Y)$ is
the ``volume'' of phase space occupied at a single time by all the points lying
on trajectories which pass, at the appropriate times, through all the cells
specified by the history $Y$, the weight function is additive in the sense that
\begin{equation} 
 YY' = 0 \hbox{ implies } W(Y+Y') = W(Y) + W(Y').
\label{e3.11} 
\end{equation}
However, this equation need not hold for a quantum system, because $W$ is
defined by the quadratic expression (\ref{e3.9}).  Indeed, in order for
(\ref{e3.11}) to hold it is necessary and sufficient that for all $Y$ and $Y'$
in $\F$,
\begin{equation} 
 YY' = 0 \hbox{ implies } \Re \lg K(Y),K(Y')\rg = 0,
\label{e3.12} 
\end{equation}
where $\Re$ denotes the real part.  We shall refer to (\ref{e3.12}) as a {\it
consistency condition}, and, in particular, as the {\it weak} consistency
condition, in contrast to the {\it strong} consistency condition:
\begin{equation} 
  YY' = 0 \hbox{ implies } \lg K(Y),K(Y')\rg = 0.
\label{e3.13} 
\end{equation}
Note that replacing $K$ by $\hat K$ everywhere in (\ref{e3.12}) or
(\ref{e3.13}) leads to an equivalent condition.

	The condition (\ref{e3.13}) is equivalent to the requirement that
\begin{equation} 
  j\neq k \hbox{ implies } \lg K(F_j),K(F_k)\rg = 0,
\label{e3.14} 
\end{equation}
for the $\{F_j\}$ in the decomposition of the identity (\ref{e2.8}). In other
words, strong consistency corresponds to requiring that the weight operators
corresponding to the minimal elements of $\F$ be orthogonal to each other.
This orthogonality requirement, which was pointed out in \cite{mce95}, is
closely related to the consistency condition employed by Gell-Mann and Hartle
\cite{gmh90,gmh93}, the vanishing of the off-diagonal elements of an
appropriate ``decoherence functional''.  To express the weak consistency
condition in similar terms requires that one replace (\ref{e3.8}) with the
inner product
\begin{equation} 
  \lg\lg A,B\rg\rg = \Re (\Tr[A\ad B]) = \lg\lg B,A\rg\rg,
\label{e3.15} 
\end{equation}
which is appropriate when the linear operators on $\H$ are thought of as
forming a {\it real} vector space (i.e., multiplication is restricted to real
scalars).  Because $\F$ consists of sums with real coefficients, (\ref{e2.9}),
a real vector space is not an unnatural object to introduce into the formalism,
even if it is somewhat unfamiliar.  Thus the counterpart of (\ref{e3.14}) in
the case of weak consistency is:
\begin{equation} 
  j\neq k \hbox{ implies } \lg\lg K(F_j),K(F_k)\rg\rg = 0.
\label{e3.14b}
\end{equation} 
The use of a weak consistency condition has the advantage that it allows a
wider class of consistent families in the quantum formalism.  However, greater
generality is not always a virtue in theoretical physics, and it remains to be
seen whether there are ``realistic'' physical situations where it is actually
helpful to employ weak rather than strong consistency.  In any case, the
formalism developed below works equally well if $\lg\, ,\,\rg$ is replaced by
$\lg\lg\, ,\,\rg\rg$, so that our use of the former can be regarded as simply a
matter of convenience of exposition.  For some further comments on the
relationship of our consistency conditions and those of Gell-Mann and Hartle,
see App.~A.

	Henceforth we shall refer to a consistent Boolean algebra of history
projectors as a {\it framework}, or {\it consistent family}, and regard it as
the appropriate quantum counterpart of the event algebra in ordinary
probability theory. Since a Boolean algebra of histories is always based upon a
decomposition of the (history) identity, as in (\ref{e2.8}), we shall say that
such a decomposition is {\it consistent} if its minimal elements satisfy
(\ref{e3.14}) or (\ref{e3.14b}), as the case may be, and will occasionally, as
a matter of convenience, refer to such a decomposition as a ``framework'',
meaning thereby the corresponding Boolean algebra which it generates.

	While the consistency condition is not essential for defining a quantum
probability, it is convenient for technical reasons, and seems to be adequate
for representing whatever can be said realistically about a {\it closed}
quantum system.  (Regarding open quantum systems, see Sec.~\ref{open}.)  Note
that while the concept of consistency properly applies to a Boolean algebra, or
a decomposition of $\IB$, an individual history $Y$ can be inconsistent in the
sense that $K(Y)$ and $K(\IB - Y)$ are not orthogonal, and hence there exists
no consistent family which contains $Y$.

	It is sometimes convenient to focus one's attention on a Boolean
algebra of histories for which the maximum element is not the identity $\IB$ on
the history space, but a smaller projector.  For example, one may be interested
in a family $\G$ of histories for which there is a fixed initial event at
$t_1$, corresponding to the projector $A$.  In this case it is rather natural
to replace (\ref{e2.8}) with
\begin{equation}
 \breve A= \sum_i G_i, \quad G_i\ad = G_i,\quad G_i G_j = \delta_{ij} G_i,
\label{e3.16}
\end{equation}
where $\breve A$ is defined as:
\begin{equation} 
 \breve A= A\od I \od I \cdots I.
\label{e3.17} 
\end{equation}
The largest projector or maximum element on the Boolean algebra of projectors
generated by the $\{G_j\}$, in analogy with (\ref{e2.9}), is $\breve A$ rather
than $\IB$.  If this algebra is consistent, which is to say the weight
operators corresponding to the different $G_i$ are mutually orthogonal, then
one can add the projector $\IB -\breve A$ to the algebra and the resulting
family, whose maximum element is now $\IB$, is easily seen to be consistent.
The same comment applies to families in which there is a fixed final event $B$,
and to those, such as in \cite{gr84}, with a fixed initial {\it and} final
event.  However, if an event $C$ at an intermediate time is held fixed, the
consistency of the family based upon the corresponding $\breve C$ is not
automatic.  Once again, it seems that for a description of closed quantum
systems, the appropriate requirement is that an acceptable framework either be
a consistent Boolean algebra whose maximum element is $\IB$, or a subalgebra of
such an algebra.

	From now on we shall adopt the following as a fundamental principle of
quantum reasoning: {\it A meaningful description of a (closed) quantum
mechanical system, including its time development, must employ a single
framework}.

		\section{Probabilities and Refinements}
\label{prob}  

	Throughout this section, and in the rest of the paper, a {\it
framework} will be understood to be a Boolean algebra of projectors on the
history space, based upon a decomposition of the identity as in (\ref{e2.8}),
and satisfying a consistency condition, either (\ref{e3.12}) or (\ref{e3.13}).
In the special case where only a single time is involved, the consistency
condition is not needed (or is automatically satisfied).

	A {\it probability distribution} $\Pr()$ on a framework $\F$ is an
assignment of a non-negative number $\Pr(Y)$ to every history $Y$ in $\F$ by
means of the formula:
\begin{equation} 
 \Pr(Y)=\sum_i \upsilon_i \Pr(F_i) = \sum_i \theta(P|F_i) \Pr(F_i),
\label{e4.1} 
\end{equation}
where the $\upsilon_i$ are defined in (\ref{e2.9}), and the probabilities
$\Pr(F_i)$ of the minimal elements are arbitrary, subject only to the
conditions:
\begin{eqnarray} 
Pr(F_i) & \geq& 0,\quad \sum_i \Pr(F_i) = 1, \label{e4.2}\\ W(F_i)&=0 &\hbox{
implies } \Pr(F_i)=0.\label{e4.3}
\end{eqnarray}
Of course, (\ref{e4.2}) are the usual conditions of any probability theory,
while (\ref{e4.3}), using the weight $W$ defined in (\ref{e3.9}), expresses the
requirement that zero probability be assigned to any history which is
dynamically impossible.  If $W(F_i)$ is zero, $\theta(P|F_i)$ is undefined, and
we set the corresponding term in the second sum in (\ref{e4.1}) equal to zero,
which is plausible in view of (\ref{e4.3}).  In addition, note that, because
the weights are additive for histories in a (consistent) framework,
(\ref{e4.3}) implies that whenever $W(Y)$ is zero, $\Pr(Y)$ vanishes.

	Apart from the requirement (\ref{e4.3}), quantum theory by itself does
not specify the probability distribution on the different histories.  Thus
these probabilities must be assigned on the basis of various data known or
assumed to be true for the quantum system of interest.  A typical example is
one in which a system is known, or assumed, to be in an initial state
$|\psi_0\rg$ at an initial time $t_0$, which would justify assigning
probabilities 1 and 0, respectively, to the projectors
\begin{equation} 
 \psi_0 = |\psi_0\rg\lg \psi_0|,\quad \tilde\psi_0 = I-\psi_0
\label{e4.4} 
\end{equation}
at the initial time.

	The process of refining a probability distribution plays an important
role in the system of quantum reasoning described in Sec.~\ref{qr} below.
We shall say that the framework $\G$ is a {\it refinement} of $\F$, and $\F$ a
{\it coarsening} of $\G$, provided $\F\subset\G$, that is, provided every
projector which appears in $\F$ also appears in $\G$.  A collection $\{\F_i\}$
of two or more frameworks is said to be {\it compatible} provided there is a
common refinement, i.e., some framework $\G$ such that $\F_i\subset\G$ for
every $i$.  If the collection is compatible, there is a smallest (coarsest)
common refinement, and we shall call this the framework {\it generated by} the
collection, or simply the {\it generated} framework.  (Note that in
constructing refinements it may be necessary to extend certain histories to
additional times by introducing an identity operator at these times, as
discussed above in Sec.~\ref{proj}.)

	Frameworks not compatible with each other are called {\it
incompatible}. Incompatibility of $\F_1$ and $\F_2$ can arise in two somewhat
different ways.  First, some of the projectors in $\F_1$ may not commute with
projectors in $\F_2$, and thus one cannot construct the Boolean algebra of
projectors needed for a common refinement.  Second, even if the common Boolean
algebra can be constructed, it may not be consistent, despite the fact that the
algebras for both $\F_1$ and $\F_2$ are consistent.

	Given a probability distribution $\Pr()$ on $\F$ and a refinement $\G$
of $\F$, we can define a probability $\Pr'()$ on $\G$ by means of the {\it
refinement rule}:
\begin{equation} 
 {\Pr}'(G) = \sum_i \theta(G|F_i) \Pr(F_i).
\label{e4.5} 
\end{equation}
Here $G$ is any projector in $\G$, and if $\Pr(F_i)$ is zero, the corresponding
term in the sum is set equal to zero, thus avoiding any problems when $\theta$
is undefined.  	Note that (\ref{e4.5}) assigns zero probability to any $G$
having zero weight, and in particular to minimal elements of $\G$ with zero
weight.  Hence $\Pr'()$ satisfies the analog of (\ref{e4.3}), and it is easily
checked that it satisfies the conditions corresponding to (\ref{e4.2}). In view
of (\ref{e4.1}) and the fact that $\G$ is a refinement of $\F$, $\Pr'(F)$ and
$\Pr(F)$ are identical for any $F\in\F$.  Consequently there is little danger
of confusion if the prime is omitted from $\Pr'()$.

	It is straightforward to show that if $\G$ is a refinement of $\F$,
$\Pr'()$ the probability on $\G$ obtained by applying the refinement rule to
$\Pr()$ on $\F$, and $\J$ a refinement of $\G$, then the same refined
probability $\Pr''()$ on $\J$ is obtained either by applying the refinement
rule to $\Pr'()$ on $\G$, or by regarding $\J$ as a refinement of $\F$, and
applying the refinement rule directly to $\Pr()$.  	Note that if $A$ is a
projector which occurs in some refinement of $\F$, then $\Pr(A)$ is the same in
any refinement of $\F$ in which $A$ occurs. This follows from noting that
$\Pr(A)$ is given by (\ref{e4.5}), with $A$ in the place of $G$, and that
$\theta(A|F_i)$ is simply a ratio of weights, and thus does not depend upon the
framework. (The same comment applies, of course, if $A$ is a member of $\F$,
and hence a member of every refinement of $\F$.)  Thus, relative to the
properties just discussed, the refinement rule is internally consistent.

	The significance of the refinement rule can best be appreciated by
considering some simple examples.  As a first example, let $\F$ be the family
whose minimal elements are the two projectors $\psi_0$ and $\tilde\psi_0$ at
the single time $t_0$, see (\ref{e4.4}), and $\G$ a refinement whose minimal
elements are of the form
\begin{equation} 
 \psi_0\od\psi_1^\alpha,\quad \tilde\psi_0\od\psi_1^\alpha,
\label{e4.6} 
\end{equation}
where the states $|\psi_1^\alpha\rg$, with $\alpha=1,2,\ldots$ form an
orthonormal basis of $\H$, and the corresponding projectors $\psi_1^\alpha$,
defined using dyads as in (\ref{e4.4}), represent properties of the quantum
system at time $t_1$.  Using the fact that
\begin{equation} 
 W(\psi_0\od\psi_1^\alpha) = |\lg \psi_1^\alpha |\psi_0\rg|^2,
\label{e4.7} 
\end{equation}
and the assumption that $\Pr(\psi_0\od I)=1$ in $\F$, one arrives at the
conclusion that
\begin{equation} 
 \Pr(\psi_0\od\psi_1^\alpha) = |\lg \psi_1^\alpha |\psi_0\rg|^2
\label{e4.8} 
\end{equation}
in $\G$, which is just the Born rule for transition probabilities.  Thus in
this example the refinement rule embodies the consequences of quantum dynamics
for the time development of the system.

	A second example involves only a single time.  Let the projector $D$ on
a subspace of dimension $d$ be a minimal element of $\F$ to which is assigned a
probability $p$.  If in the refinement $\G$ of $\F$ one has two minimal
elements $D_1$ and $D_2$, projectors onto subspaces of dimension $d_1$ and
$d_2$, whose sum is $D$, then in the refined probability $\Pr'()$, $D_1$ is
assigned a probability $pd_1/d$ and $D_2$ a probability $pd_2/d$.  That is to
say, the original probability is split up according to the sizes of the
respective subspaces.  While in this example the refinement rule is not a
consequence of the dynamical laws of quantum theory, it is at least not
inconsistent with them.

	The following result on conditional probabilities is sometimes useful.
Let $D$ be a minimal element of a framework $\D$ having positive weight, and
assign to $\D$ the probability
\begin{equation} 
 \Pr(D)=1,\quad \Pr(\IB-D)=0.
\label{e4.9} 
\end{equation}
Let $\E$ be a refinement of $\D$, and $E$ some element of $\E$ with positive
weight such that
\begin{equation} 
 ED = E.
\label{e4.10} 
\end{equation}
Then for $E'$ any element of $\E$,
\begin{equation} 
 \Pr(E'|E) = \ta(E'|E).
\label{e4.11}  
\end{equation}
We omit the derivation, which is straightforward.  Note that it is essential
that $D$ be a minimal element of $\D$, and that (\ref{e4.10}) be satisfied; it
is easy to construct examples violating one or the other of these conditions
for which (\ref{e4.11}) does not hold.

		\section{Quantum Reasoning}
\label{qr}

	The type of quantum reasoning we shall focus on in this section is that
in which one starts with some information about a system, known or assumed to
be true, and from these {\it initial data} tries to reach valid {\it
conclusions} which will be true if the initial data are correct.  As is usual
in logical systems, the rules of reasoning do not by themselves certify the
correctness of the initial data; they merely serve to define a valid process of
inference.  Note that the term ``initial'' refers to the fact that these data
represent the beginning of a logical argument, and has nothing to do with the
temporal order of the data and conclusions in terms of the history of the
quantum system.  Thus the conclusions of the argument may well refer to a point
in time prior to that of the initial data.

	Since quantum mechanics is a stochastic theory, the initial data and
the final conclusions will in general be expressed in the form of
probabilities, and the rules of reasoning are rules for deducing probabilities
from probabilities.  In this context, ``logical rules'' for deducing true
conclusions from true premises refer to limiting cases in which certain
probabilities are 1 (true) or 0 (false).  Since probabilities in ordinary
probability theory always refer to some sample space, we must embed quantum
probabilities referring to properties or the time development of a quantum
system in an appropriate framework.  Both the initial data and the final
conclusions of a quantum argument should be thought of as labeled by the
corresponding frameworks.  Likewise, the truth or falsity of a quantum
proposition, and more generally its probability, is relative to the framework
in which it occurs.

	As long as only a single framework is under discussion, the rules of
quantum reasoning are the usual rules for manipulating probabilities.  In
particular, if the initial data is given as a probability distribution $\Pr()$
on a framework $\D$, we can immediately say that a proposition represented by a
projector $D$ in $\D$ with $\Pr(D)=1$ is true (in the framework $\D$ and
assuming the validity of the initial data), whereas if $\Pr(D)=0$, the
proposition is false (with the same qualifications).  Given a framework $\D$,
there are certain propositions for which the probability is 1 for any
probability distribution satisfying the rules (\ref{e4.2}) and (\ref{e4.3}),
and we call these {\it tautologies}; their negations are {\it contradictions}.
For example, given any $D\in \D$, the proposition ``$D$ or not $D$'', which
maps onto the projector $D\lor (\IB-D)=\IB$, is always true, whereas any
history in $\D$ which has zero weight, meaning that it violates the dynamical
laws, is always false.

	Arguments which employ only a single framework are too restrictive to
be of much use in quantum reasoning.  Hence we add, as a fundamental principle,
the following {\it refinement rule\/}: if a probability distribution $\Pr()$ is
given for a framework $\F$, and $\G$ is a refinement of $\F$, then one can
infer the probability distribution $\Pr'()$ on $\G$ given by the refinement
rule introduced in Sec.~\ref{prob}, see (\ref{e4.5}).  As noted in
Sec.~\ref{prob}, the refinement rule embodies all the dynamical consequences of
quantum theory.  Replacing $\Pr'()$ by $\Pr()$ will generally cause no
confusion, because the two are identical on $\F$.

	Thus the general scheme for quantum reasoning is the following.  One
begins with data in the form of a probability distribution $\Pr()$ on a
framework $\D$, calculates the refined probability distribution on a refinement
$\E$ of $\D$, and applies the standard probability calculus to the result.
Note that the internal consistency of the refinement rule of Sec.~\ref{prob}
has the following important consequence: If a history $A$ occurs in some
refinement of $\D$, then $\Pr(A)$ is the same in any refinement of $\D$ in
which $A$ occurs. In particular, it is impossible to deduce from the same
initial data that some proposition $A$ is both true (probability 1) and false
(probability 0).  In this sense the scheme of quantum reasoning employed here
is internally consistent.

	Even in the case of ``complete ignorance'', that is to say, in the
absence of any initial data, this scheme can generate useful results. Consider
the trivial framework $\D = \{0,\IB\}$ for which the only probability
assignment consistent with (\ref{e4.2}) and (\ref{e4.3}) is $\Pr(\IB)=1$.  Let
$\E$ be any framework which uses the same Hilbert space as $\D$, and which is
therefore a refinement of $\D$.  For any $E'$ and $E$ in $\E$ with $W(E)>0$,
(\ref{e4.11}) applies, so that a logical consequence of complete ignorance is:
\begin{equation} 
 \Pr(E'|E) = \ta(E'|E).
\label{e5.1} 
\end{equation}

	For example, if we apply (\ref{e5.1}) to the case where $\E$ is the
framework consisting of the elements in (\ref{e4.6}), one consequence is:
\begin{equation} 
 \Pr(\psi_1^\alpha|\psi_0) = |\lg \psi_1^\alpha |\psi_0\rg|^2.
\label{e5.2} 
\end{equation}
Hence while we cannot, in the absence of initial data, say what the initial
state is, we can nevertheless assert that {\it if\/} the initial state is
$\psi_0$ at $t_0$, {\it then\/} at $t_1$ the probability of $\psi_1^\alpha$ is
given by (\ref{e5.2}). Thus, even complete ignorance allows us to deduce the
Born formula as a {\it conditional} probability.

	In the case in which some (nontrivial) initial data are given, perhaps
consisting of separate pieces of information associated with different
frameworks, these must first be combined into a single probability distribution
associated with a single framework before the process of refinement can begin.
For example, the data may consist of a collection of pairs $\{(\D_i,D_i)\}$,
where $D_i$ is known or assumed to be true in framework $\D_i$.  If the
$\{\D_i\}$ are incompatible frameworks, the initial data must be rejected as
mutually incompatible; they cannot all apply to the same physical system.  If
they are compatible, let $\D$ be the framework they generate, and let
\begin{equation} 
 D=D_1 D_2 D_3 \cdots
\label{e5.3} 
\end{equation}
be the projector corresponding to the simultaneous truth of the different
$D_i$. Then we assign probability 1 to $D$ and 0 to its complement $\IB-D$ in
the framework $\D$.  (Of course, this probability assignment is impossible if
$W(D)=0$, which indicates inconsistency in the initial data.) Note that if $D$
is a minimal element of $\D$, then conditional probabilities are given directly
in terms of the $\ta$ function, (\ref{e4.11}), for any $E$ satisfying
(\ref{e4.10}).

	Of course, in general the initial data may be given not in the form of
certain projectors known (or assumed) to be true, but instead as probabilities
in different frameworks. If the frameworks are incompatible, the data, of
course, must be rejected as mutually incompatible.  If the frameworks are
compatible, the data must somehow be used to generate a probability
distribution on the generated framework $\D$.  We shall not discuss this
process, except to note that because it can be carried out in the single
framework $\D$, whatever methods are applicable for the corresponding case of
``classical probabilities'' can also be applied to the quantum problem.

	The requirement that the initial data be embodied in a single framework
is just a particular example of the general principle already stated at the end
of Sec.~\ref{wc}: quantum descriptions, and thus quantum reasoning referring to
such descriptions, must employ a single framework.  This requirement is not at
all arbitrary when one remembers that probabilities in probability theory only
have a meaning relative to some sample space or algebra of events, and that the
quantum framework is playing the role of this algebra.  Probabilities in
classical statistical mechanics satisfy precisely the same requirement, where
it is totally uninteresting because there is never any problem combining
information of various sorts into a common description using, say, a single
coarse graining of the phase space (or a family of coarse grainings indexed by
the time).  What distinguishes quantum from classical reasoning is the presence
in the former, but not in the latter, of incompatible frameworks.  Thus the
rules governing incompatible frameworks are necessarily part of the foundations
of quantum theory itself.

	Note that the system of reasoning employed here does {\it not} allow a
``coarsening rule'' in which, if $\F$ is a refinement of $\E$, and a
probability distribution $\Pr()$ is given on $\F$, one can from this deduce a
probability distribution $\Pr^*()$ on $\E$ which is simply the restriction of
$\Pr()$ to $\E$, i.e.,
\begin{equation} 
 E\in \E:\quad{\Pr}^*(E) = \Pr(E).
\label{e5.4}
\end{equation}
The reason such a coarsening rule is not allowed is that if it is combined with
the refinement rule, the result is a system of reasoning which is internally
inconsistent.  For example, if we start with the probability distribution
$\Pr()$ on $\F$, define $\Pr^*$ on $\E$ by means of (\ref{e5.3}), and then
apply the refinement rule to $\Pr^*$ in order to derive a probability
${\Pr^*}'$ on $\F$, the latter will in general not coincide with the original
$\Pr()$.  Worse than this, there are cases in which successive applications of
coarsening and refinement to different quantum frameworks can lead to
contradictions: starting with $\Pr(A)=1$ in one framework one can eventually
deduce $\Pr(A)=0$ in the same framework.  To be sure, it is the combination of
coarsening and refinement which gives rise to inconsistencies, and the system
of reasoning would be valid if only the coarsening rule were permitted.
However, such a system would not be very useful.  And, indeed, there is a sense
in which a coarsening rule is also not really needed.  If $\F$ is a refinement
of $\E$, and a probability distribution is given on $\F$, then it already
assigns a probability to every projector $E$ in $\E$, in the sense that $E$ is
already an element of $\F$.  But once again this serves to emphasize the fact
that the question of which sample space one is using, while usually a trivial
and uninteresting question in classical physics, is of utmost importance in
quantum theory.

	One way of viewing the difference between quantum and classical
reasoning is that whereas in both cases the validity of a conclusion depends
upon the data from which it was derived, in the classical case one can forget
about the data once the conclusion has been obtained, and no contradiction will
arise when this conclusion is inserted as the premise of another argument.  In
the quantum case, it is safe to forget the original data {\it as a probability
distribution}, but the fact that the data were embodied in a particular {\it
framework} cannot be ignored: the conclusion must be expressed relative to a
framework, and since that framework is either identical to, or has been
obtained by refinement of the one containing the initial data, the ``framework
aspect'' of the initial data has not been forgotten.  The same is true, of
course, in the classical case, but the framework can safely be ignored, because
classical physics does not employ incompatible frameworks.

	Another way in which quantum reasoning is distinctly different from its
classical counterpart is that from the {\it same data} it is possible to draw
{\it different conclusions} in {\it mutually incompatible frameworks}. Because
the frameworks are incompatible, the conclusions cannot be combined, a
situation which is bizarre from the perspective of classical physics, where it
never arises.  See the examples below, and the discussion in Sec.~\ref{incom}.

			\section{Examples}
\label{exams}

		\subsection{Spin Half Particle}
\label{spin}

	As a first example, consider a spin one-half particle, for which the
Hilbert space is two dimensional, and a framework $\Z$ corresponding to a
decomposition of the identity:
\begin{equation} 
 I=Z^+ + Z^-, \quad Z^\pm = |Z^\pm \rg \lg Z^\pm|,
\label{e6.1} 
\end{equation}
where $|Z^+\rg$ and $|Z^-\rg$ are the states in which $S_z$ has the values
$+1/2$ and $-1/2$, respectively, in units of $\hbar$.  Within this framework,
the statement ``$S_z=1/2$ or $S_z=-1/2$'' is a tautology because it corresponds
to the projector $I$, see (\ref{e2.5}), which has probability 1 no matter what
probability distribution is employed.  Also, if $S_z=1/2$ is true (probability
1), then $S_z=-1/2$ is false (probability 0), because $\Pr(Z^+)+\Pr(Z^-)$ is
always equal to one.

	Of course, we come to precisely the same type of conclusion if, instead
of $\Z$, we use the framework $\X$ corresponding to:
\begin{equation} 
 I=X^+ + X^-; \quad X^\pm = |X^\pm \rg \lg X^\pm|,
\label{e6.2} 
\end{equation} 
where
\begin{equation} 
 |X^+\rg = (|Z^+\rg + |Z^-\rg)/\st,\quad |X^-\rg = (|Z^+\rg - |Z^-\rg)/\st
\label{e6.3} 
\end{equation}
are states in which $S_x$ is $+1/2$ or $-1/2$.  However, the frameworks $\Z$
and $\X$ are clearly incompatible because the projectors $X^\pm$ do not commute
with $Z^\pm$.  Therefore, whereas $S_z=1/2$ is a meaningful statement, which
may be true or false within the framework $\Z$, it makes no sense within the
framework $\X$, and, similarly, $S_x=1/2$ is meaningless within the framework
$\Z$.  Consequently, ``$S_z=1/2$ {\it and\/} $S_x=1/2$'' is a meaningless
statement within quantum mechanics interpreted as a stochastic theory, because
a meaningful description of a quantum system must belong to some framework, and
there is no framework which contains both $S_z=1/2$ and $S_x=1/2$ at the same
instant of time.

	A hint that ``$S_z=1/2$ {\it and\/} $S_x=1/2$'' is meaningless can also
be found in elementary textbooks, where the student is told that there is no
way of simultaneously {\it measuring\/} both $S_z$ and $S_z$, because
attempting to measure one component will disturb the other in an uncontrolled
way.  While this is certainly true, one should note that the fundamental reason
no simultaneous measurement of both quantities is possible is that there is
nothing to be measured: the simultaneous values do not exist.  Even very good
experimentalists cannot measure what is not there; indeed, this inability helps
to distinguish them from their less talented colleagues.  We return to the
topic of measurement in Sec.~\ref{measure} below.

	As an application of the refinement rule of Sec.~\ref{qr}, we can start
with ``complete ignorance'', expressed by assigning probability 1 to $I$ in the
framework $\D=\{0,I\}$, and refine this to a probability on $\Z$.  The result
is:
\begin{equation} 
 \Pr(Z^+)=1/2=\Pr(Z^-),
\label{e6.4} 
\end{equation}
that is, the particle is unpolarized.  Were we instead to use $\X$ as a
refinement of $\D=\{0,I\}$, the conclusion would be
\begin{equation} 
 \Pr(X^+)=1/2=\Pr(X^-).
\label{e6.5} 
\end{equation}	
Thus we have a simple example of how quantum reasoning starting from a
particular datum (in this case the rather trivial $\Pr(I)=1$) can reach two
different conclusions in two different frameworks. Each conclusion is correct
by itself, in the sense that it could be checked by experimental measurement,
but the conclusions cannot be combined into a common description of a single
quantum system.

		\subsection{Harmonic Oscillator}
\label{harmony}

	The intuitive or ``physical'' meaning of a projector on a subspace of
the quantum Hilbert space depends to some extent on the framework in which this
projector is embedded, as illustrated by the following example.

	Let $|n\rg$ with energy $(n+1/2)\hbar\omega$ denote the $n'$th energy
eigenstate of a one-dimensional oscillator.  (In order to have a
finite-dimensional Hilbert space, we must introduce an upper bound for $n$; say
$n<10^{80}$.)  Throughout the following discussion it will be convenient to
assume that the energy is expressed in units of $\hbar\omega$, or,
equivalently, $\hbar\omega=1.$

Define the projectors
\begin{equation} 
 B_n=|n\rg\lg n|, \quad P= B_1+B_2,\quad \Pt = I-P.
\label{e6.6} 
\end{equation}
	In any framework which contains it, $P$ can be interpreted to mean that
``the energy is less than $2$'', but in general it is {\it not} correct to
think of $P$ as meaning ``the energy is $1/2$ or $3/2$''. The latter is a
correct interpretation of $P$ in the framework based on
\begin{equation} 
 I=B_0 + B_1 + \Pt,
\label{e6.7} 
\end{equation}
because the projectors $B_0$ and $B_1$ can be interpreted as saying that the
energy is $1/2$ and $3/2$, respectively, and $P$ is their sum; see
(\ref{e2.5}).  However, it is totally incorrect to interpret $P$ to mean ``the
energy is $1/2$ or $3/2$'' when $P$ is an element in the framework based on
\begin{equation} 
 I=C^+ + C^- + \Pt,
\label{e6.8} 
\end{equation}
where $C^+$ and $C^-$ are projectors onto the states
\begin{equation} 
 |+\rg = (|0\rg + |1\rg)/\st,\quad |-\rg = (|0\rg - |1\rg)/\st.
\label{e6.9} 
\end{equation}
Because $C^+$ and $C^-$ do not commute with $B_0$ and $B_1$, the assertion that
``the energy is $1/2$'' makes no sense if we use (\ref{e6.8}), and the same is
true of ``the energy is $3/2$''.  Combining them with ``or'' does not help the
situation unless one agrees that ``the energy is $1/2$ or $3/2$'' is a sort of
shorthand for the correct statement that ``the energy is not greater than
$3/2$''.  And since even the latter can easily be misinterpreted, it is perhaps
best to use the projector $P$ itself, as defined in (\ref{e6.6}), rather than
an ambiguous English phrase, if one wants to be very careful and avoid all
misunderstanding.

	The meaning of $P$ in the smallest framework which contains it, the one
based upon
\begin{equation} 
 I=P + \Pt,
\label{e6.10} 
\end{equation}
involves an additional subtlety.  Since neither $B_0$ nor $B_1$ are part of
this framework, it is, at least formally, incorrect to say that within this
framework $P$ means ``the energy is $1/2$ or $3/2$''.  On the other hand, the
(assumed) truth of $P$ in (\ref{e6.10}) corresponds to $\Pr(P)=1$, and since
(\ref{e6.7}) is a refinement of (\ref{e6.10}), the refinement rule allows us to
conclude that the probability of $B_0+B_1$ in (\ref{e6.7}) is also equal to
one, and therefore ``$B_0$ or $B_1$'' is true in the framework (\ref{e6.7}).
And since, at least in informal usage, the ``meaning'' of a physical statement
includes various logical consequences which the physicist regards as more or
less intuitively obvious, part of the informal meaning or ``aura'' of $P$ in
the framework (\ref{e6.10}) is ``$B_0$ or $B_1$''.  However, because of the
possibility of making alternative logical deductions from the truth of $P$,
such as ``$C^+$ or $C^-$'', the best policy, if one wants to be precise, is to
pay attention to the framework, and say that the truth of $P$ in (\ref{e6.10})
means that ``the energy is $1/2$ or $3/2$ {\it in the framework based upon}
(\ref{e6.7}).''  To be sure, in informal discourse one might omit the final
qualification on the grounds that the phrase ``the energy is $1/2$ or $3/2$''
itself singles out the appropriate framework.  The point, in any case, is that
quantum descriptions necessarily take place inside frameworks, and clear
thinking requires that one be able to identify which framework is being used at
any particular point in an argument.

	As another example of a possible pitfall, suppose that we know that the
energy is $5/2$.  Can we conclude from this that the energy is {\it not} equal
to $1/2$?  There is an almost unavoidable temptation to say that the second
statement is an immediate consequence of the first, but in fact it is or is not
depending upon the framework one is using. To say that the energy is $5/2$
means that we are employing a framework which includes $B_2$ as one of its
elements.  If this framework also includes $B_0$, the fact that $B_0$ is false
(probability 0) follows at once from the assumption that $B_1$ is true
(probability 1), by an elementary argument of probability theory, so that,
indeed, the energy is not equal to $1/2$. If the framework does not include
$B_0$, but has some refinement which does include $B_0$, we can again conclude
that {\it within this refined framework\/}---which, note, is not the original
framework---the energy is not equal to $1/2$.  However, if the original
framework is incompatible with $B_0$ (e.g., it might contain $C^+$), then the
fact that the energy is $5/2$ does {\it not} imply that the energy is not equal
to $1/2$!  Ignoring differences between different frameworks quickly leads to
paradoxes, as in the example in Sec.~\ref{vaidman} below.

		\subsection{Measurement of Spin}
\label{measure}

	Textbook discussions of quantum measurement suffer from two distinct
but related ``measurement problems''.  The first is that the use of unitary
time development can result in MQS (macroscopic quantum superposition) or
``Schr\"odinger's cat'' states, which must then somehow be explained away in a
manner which has been justly criticized by Bell \cite{bl90}.  The second is
that many measurements of properties of quantum particles, such as energy or
momentum, when actually carried out in the laboratory result in large changes
in the measured property.  Since one is generally interested in the property of
the particle before its interaction with the measurement apparatus, the
well-known von Neumann ``collapse'' description of the measurement is
unsatisfactory (quite aside from the never-ending debate about what such a
``collapse'' really means).  The system of quantum reasoning developed in
Sec.~\ref{qr} resolves both problems through the use of appropriate frameworks,
as illustrated in the following discussion of the measurement of the spin of a
spin-half particle.

	The particle and the measuring apparatus should be thought of as a
single closed quantum system, with Hilbert space
\begin{equation} 
 \H = \SS\ot\A.
\label{e6.11} 
\end{equation}
Here $\SS$ is the two-dimensional spin space for the spin-half particle, and
$\A$ is the Hilbert space for all the remaining degrees of freedom: the
particles constituting the apparatus, and the center of mass of the spin-half
particle.  We consider histories involving three times, $t_0 < t_1 < t_2$, and
suppose that the relevant unitary time development, indicated by $\mt$, has the
form:
\begin{equation} 
	\begin{array}{c} 	|Z^+A\rg \mt |Z^+A'\rg \mt |P^+\rg \\
|Z^-A\rg \mt |Z^-A'\rg \mt |P^-\rg \\ 	\end{array}
\label{e6.12} 
\end{equation}
where $|Z^+\rg$ and $|Z^-\rg$ are the spin states for $S_z$ equal to $\pm 1/2$,
as in (\ref{e6.1}), $|A\rg$ is a state on $\A$ at time $t_0$ in which the
particle is traveling towards the apparatus, and the apparatus is ready for the
measurement, $|A'\rg$ is the corresponding state at $t_1$, with the particle
closer to, but still not at the apparatus, and $|P^+\rg$ and $|P^-\rg$ are
states on $\H$ at $t_2$, after the measurement is complete, which correspond to
the apparatus indicating, through the position of a pointer, the results of
measuring $S_z$ for the particle.  Note that the spin state of the particle at
$t_2$ is included in $|P^+\rg$ and $|P^-\rg$, and we do {\it not} assume that
it remains unchanged during the measuring process.  Such a description using
only pure states is oversimplified, but we will later indicate how essentially
the same results come out of a more realistic discussion.

	To keep the notation from becoming unwieldy, we use the following
conventions.  A letter outside a ket indicates the dyad for the corresponding
projector; e.g., $A$ stands for $|A\rg\lg A|$.  Next, we make no distinction in
notation between $A$ as a projector on $\A$ and as the projector $I\ot A$ on
$\SS\ot \A$; similarly, $Z^+$ stands both for the projector on $\SS$ and for
$Z^+\ot I$ on $\H$.  Finally, projectors on the history space $\HB$ carry
subscripts which indicate the time, as in the following examples:
\begin{equation} 
 A_0 = A\od I \od I, \quad P^+_2 = I \od I \od P^+.
\label{e6.13} 
\end{equation}

	We first consider a framework associated with the decomposition
\begin{equation} 
 \IB = \At_0 + \{Z^+_0A_0 + Z^-_0A_0\}\{P^+_2 + P^-_2 + P^*_2 \},
\label{e6.14} 
\end{equation}
containing seven minimal elements, of the identity on $\HB$, where
\begin{equation} 
 \At = I-A,\quad P^*= I-(P^+ + P^-).
\label{e6.15} 
\end{equation}
The family generated by (\ref{e6.14}) is easily shown to be consistent, and the
following weights are a consequence of (\ref{e6.12}):
\begin{equation}
	\begin{array}{c} W(Z^+_0A_0\cdot P^+_2) = 1 = W(Z^-_0A_0\cdot P^-_2),
\\ W(Z^-_0A_0\cdot P^+_2) = 0 = W(Z^+_0A_0\cdot P^-_2). \\ 	\end{array}
\label{e6.16} 
\end{equation}
In addition, weights of histories which include both $A_0$ and $P^*_2$ vanish.
Note that the weights are additive, so that, for example,
\begin{equation} 
 W(A_0\cdot P^+_2) = W(Z^+_0A_0\cdot P^+_2) + W(Z^-_0A_0\cdot P^+_2) = 1.
\label{e6.17} 
\end{equation}
If we assume that the initial data correspond either to ``complete ignorance'',
see the remarks preceding (\ref{e5.1}), or to probability 1 for $A_0$ in the
framework corresponding to $\IB=A_0+\At_0$, see (\ref{e4.9}), we can equate
probabilities which include $A_0$ as a condition with the corresponding $\ta$
functions, (\ref{e4.11}), and the latter can be computed using (\ref{e3.10}).
The results include:
\begin{eqnarray} 
 \Pr(P^+_2|Z^+_0 A_0) &=& 1, \quad \Pr(P^-_2|Z^+_0 A_0) = 0, \label{e6.18} \\
\Pr(P^+_2|A_0) &=& 1/2 = \Pr(P^-_2|A_0), \label{e6.19}\\ \Pr(Z^+_0 | P^+_2 A_0)
&=&1,\quad \Pr(Z^-_0 | P^+_2 A_0)=0.\label{e6.20}
\end{eqnarray}
The probabilities in (\ref{e6.18}) are certainly what we would expect: if at
$t_0$ we have $S_z=1/2$, then at $t_2$ the apparatus pointer will surely be in
state $P^+$ and not in state $P^-$.  On the other hand, if we are ignorant of
$S_z$ at $t_0$, the results in (\ref{e6.19}) are those appropriate for an
unpolarized particle.  Equally reasonable is the result (\ref{e6.20}), which
tells us that if at $t_2$ the pointer is at $P^+$, the spin of the particle at
$t_0$ was given by $S_z=1/2$, not $S_z=-1/2$; that is, the measurement reveals
a property which the particle had before the measurement took place.

	Next consider, as an alternative to (\ref{e6.14}), the framework based
upon:
\begin{equation} 
 \IB = \At_0 + \{X^+_0A_0 + X^-_0A_0\}\{P^+_2 + P^-_2 + P^*_2 \},
\label{e6.21} 
\end{equation}
where $X^+$ and $X^-$ are projectors associated with $S_x=\pm 1/2$, see
(\ref{e6.3}). It is straightforward to check consistency and calculate the
weights:
\begin{equation}
	\begin{array}{c} W(X^+_0A_0\cdot P^+_2) = 1/2 = W(X^-_0A_0\cdot P^-_2),
\\ W(X^-_0A_0\cdot P^+_2) = 1/2 = W(X^+_0A_0\cdot P^-_2). \\ 	\end{array}
\label{e6.22} 
\end{equation}
Once again, weights of histories which include both $A_0$ and $P^*_2$ vanish.
With the same assumptions as before (ignorance, or $A_0$ at $t_0$), we obtain:
\begin{eqnarray} 
 \Pr(P^+_2|X^+_0 A_0) &=& 1/2, \quad \Pr(P^-_2|X^+_0 A_0) = 1/2, \label{e6.23}
\\ \Pr(X^+_0 | P^+_2 A_0) &=&1/2,\quad \Pr(X^-_0 | P^+_2 A_0)=1/2.\label{e6.24}
\end{eqnarray}
In addition, the probabilities in (\ref{e6.19}) are the same in the new
framework as in the old, which is not surprising, since they make no reference
to $S_z$ or $S_x$ at $t_0$.

	Everyone agrees that (\ref{e6.23}), assigning equal probability to the
pointer states $P^+$ and $P^-$ if at $t_0$ the spin state is $S_x=1/2$, is the
right answer.  What is interesting is that, with the formalism used here, the
right answer emerges without having to make the slightest reference to an MQS
state, and thus there is no need to make excuses of the ``for all practical
purposes'' type in order to get rid of it.  How have we evaded the problem of
Schr\"odinger's cat?

	The answer is quite simple: there is no MQS state at $t_2$ in the
decomposition of the identity (\ref{e6.21}), and therefore there is no
reference to it in any of the probabilities.  To be sure, we could have
investigated an alternative framework based upon
\begin{equation} 
 \IB = \At_0 + \{X^+_0A_0 + X^-_0A_0\}\{Q^+_2 + Q^-_2 + P^*_2 \},
\label{e6.25} 
\end{equation}
where
\begin{equation} 
 |Q^+\rg = (|P^+\rg + |P^-\rg)/\st, \quad |Q^-\rg = (|P^+\rg - |P^-\rg)/\st,
\label{e6.26} 
\end{equation}
are MQS states.  Using this framework one can calculate, for example,
\begin{equation} 
 \Pr(Q^+_2|X^+_0 A_0)=1, \quad \Pr(Q^-_2|X^+_0 A_0)=0.
\label{e6.27} 
\end{equation}
Note that there is no contradiction between (\ref{e6.27}) and (\ref{e6.23}),
because they have been obtained using mutually incompatible frameworks. Here is
another illustration of the fact that quantum reasoning based upon the same
data will lead to different conclusions, depending upon which framework is
employed.  However, conclusions from incompatible frameworks cannot be
combined, and the overall consistency of the reasoning scheme is guaranteed,
see the discussion in Sec.~\ref{qr}, by the fact that only refinements of
frameworks are permitted and coarsening is not allowed.

	Also note that the framework generated by
\begin{equation} 
 \IB = \At_0 + \{Z^+_0A_0 + Z^-_0A_0\}\{Q^+_2 + Q^-_2 + P^*_2 \}
\label{e6.28} 
\end{equation}
is just as acceptable as that based upon (\ref{e6.14}), and one can perfectly
well calculate various probabilities, such as $\Pr(Q^+_2|Z^+_0 A_0)$, by means
of it.  In this case the initial state corresponds to a definite value of
$S_z$, and yet the states at $t_2$ are MQS states! What this shows is that the
real ``measurement problem'' is not the presence of MQS states in certain
frameworks; instead, it comes about because one is attempting to address a
particular question---$P^+$ or $P^-$?---by means of a framework in which this
question has no meaning, and hence no answer.  Trying to claim that the
projector $Q^+$ is somehow equivalent to the density matrix $(P^+ + P^-)/2$ for
all practical (or any other) purposes is simply making a second mistake in
order to correct the results of a more fundamental mistake: using the wrong
framework for discussing pointer positions.  A major advantage of treating
quantum mechanics as a stochastic theory from the outset, rather than adding a
probabilistic interpretation as some sort of addendum, is that it frees one
from having to think that a quantum system ``must'' develop unitarily in time,
and then being forced to make a thousand excuses when the corresponding
framework is incompatible with the world of everyday experience.

	While the framework based upon (\ref{e6.21}) solves the first
measurement problem in the case of a particle which at $t_0$ has $S_x=1/2$, and
is traveling towards an apparatus which will measure $S_z$, it does not solve
the second measurement problem, that of showing that if the apparatus is in the
$P^+$ state at $t_2$, then the particle actually was in the state $S_z=1/2$
before the measurement.  Indeed, we cannot even introduce the projectors
$Z^+_0$ and $Z^-_0$ into the family based on (\ref{e6.21}), because they do not
commute with $X^+_0$ and $X^-_0$.  However, nothing prevents us from
introducing them at the later time $t_1$, and considering the following
refinement of (\ref{e6.21}):
\begin{equation} 
 \IB = \At_0 + \{X^+_0A_0 + X^-_0A_0\} 	\{Z^+_1 + Z^-_1\}\{P^+_2 + P^-_2 +
P^*_2 \}.
\label{e6.29} 
\end{equation}  
After checking consistency, one can calculate the following weights:
\begin{equation}
	\begin{array}{c} W(X^+_0A_0\cdot Z^+_1\cdot P^+_2) = 1/2 =
W(X^-_0A_0\cdot Z^-_1 \cdot P^-_2), \\ W(X^-_0A_0 \cdot Z^+_1\cdot P^+_2) = 1/2
= 	W(X^+_0A_0\cdot Z^-_1\cdot P^-_2). \\ 	\end{array}
\label{e6.30} 
\end{equation}
In addition, all the weights with $Z^+_1$ followed by $P^-_2$, or $Z^-_1$
followed by $P^+_2$, vanish.  Conditional probabilities can then be computed in
the same way as before, with (among others) the following results:
\begin{equation} 
  \Pr(Z^+_1 | P^+_2 X^+_0 A_0) =1,\quad \Pr(Z^-_1 | P^+_2X^+_0 A_0)=0.
\label{e6.31} 
\end{equation}
That is, given the initial condition $X^+A$ at $t_0$, and the pointer state
$P^+$ at $t_2$, one can be certain that $S_z$ was equal to $1/2$ and not $-1/2$
at the time $t_1$ before the measurement took place.

	It may seem odd that we can discuss a history in which the particle has
$S_x=1/2$ at $t_0$ and $S_z=1/2$ at $t_1$ in the absence of a magnetic field
which could re-orient its spin.  To see why there is no inconsistency in this,
note that whereas in the two-dimensional Hilbert space $\SS$ appropriate for a
spin half particle at a single time there is no way to describe a particle
which simultaneously has $S_x=1/2$ and $S_z=1/2$, the same is not true in the
history space $\SS\od\SS$ for the two times $t_0$ and $t_1$, which is four
dimensional, and hence analogous to the tensor product space appropriate for
describing two (non-identical) spin-half particles.  The fact that the
``incompatible'' spin states occur at different times is the reason that all
thirteen projectors on the right side of (\ref{e6.29}) commute with one
another.  To be sure, spin directions cannot be chosen arbitrarily at a
sequence of different times without violating the consistency conditions, but
in the case of (\ref{e6.29}) these conditions are satisfied.  	It is also
useful to remember that were we applying classical mechanics to a spinning
body, there would be no problem in ascribing a definite value to the $x$
component of its angular momentum at one time, and to the $z$ component of its
angular momentum at a later time.  That this is (sometimes) possible in the
quantum case should, therefore, not be too surprising, as long as one can make
sense of this in the appropriate Hilbert space (of histories).

	In place of (\ref{e6.29}) we could, of course, use a framework
\begin{equation} 
 \IB = \At_0 + \{X^+_0A_0 + X^-_0A_0\} 	\{X^+_1 + X^-_1\}\{P^+_2 + P^-_2 +
P^*_2 \}.
\label{e6.32} 
\end{equation}
appropriate for discussing the value of $S_x$ at $t_1$, and from it deduce the
results
\begin{equation} 
  \Pr(X^+_1 | P^+_2 X^+_0 A_0) =1,\quad \Pr(X^-_1 | P^+_2X^+_0 A_0)=0,
\label{e6.33} 
\end{equation}
in place of (\ref{e6.31}).  Note, however, that (\ref{e6.32}) and (\ref{e6.29})
are incompatible frameworks, so that one cannot combine (\ref{e6.31}) and
(\ref{e6.33}) in any way.

	What is the physical significance of two conclusions, (\ref{e6.31}) and
(\ref{e6.33}), based upon the same initial data, which are incompatible because
the deductions were carried out using incompatible frameworks?  One way of
thinking about this is to note that (\ref{e6.31}) could be verified by an
appropriate idealized measurement which would determine the value of $S_z$ at
$t_1$ without perturbing it, and similarly (\ref{e6.33}) could be checked by a
measurement of $S_x$ at $t_1$ which did not perturb that quantity \cite{gr84n}.
However, carrying out both measurements at the same time is not possible.

	In summary, the solution of quantum measurement problems, which has
hitherto led to a never-ending debate, consists in choosing an appropriate
framework.  If one wants to find out what the predictions of quantum theory are
for the position of a pointer at the end of a measurement, it is necessary (and
sufficient) to use a framework containing projectors corresponding to the
possible positions.  If one wants to know how the pointer position is
correlated with the corresponding property of the particle before the
measurement took place, it is necessary (and sufficient) to employ a framework
containing projectors corresponding to this property at the time in question.
While these criteria do not define the framework uniquely, they suffice,
because the consistency of the quantum reasoning process as discussed in
Sec.~\ref{qr} ensures that the same answers will be obtained in any framework
in which one can ask the same questions.

	As noted above, a description of the measurement process based solely
upon pure states, as in (\ref{e6.12}) is not very realistic.  It would be more
reasonable to replace the one-dimensional projectors $A$, $A'$, with projectors
of very high dimension (corresponding to a macroscopic entropy). This can,
indeed, be done without changing the main conclusions.  Thus let $A$ be a
projector onto a subspace of $\A$ of arbitrarily large (but finite) dimension
spanned by an orthonormal basis $|a_j\rg$, and replace the unitary time
evolution (\ref{e6.12}) with
\begin{equation} 
	\begin{array}{c} 	|Z^+a_j\rg \mt |Z^+a'_j\rg \mt |b^+_j\rg, \\
|Z^-a_j\rg \mt |Z^-a'_j\rg \mt |b^-_j\rg, \\ 	\end{array}
\label{e6.34} 
\end{equation}
where the $|a'_j\rg$ are, again, a collection of orthonormal states in $\A$,
while the $|b^\pm_j\rg$ are orthonormal states on $\H$, the exact nature of
which is of no particular interest aside from the fact that they satisfy
(\ref{e6.35}) below. Note in particular that nothing is said about the spin of
the particle at $t_2$, as that is entirely irrelevant for the measuring
process.  Next we assume that $P^+$ and $P^-$ are projectors onto enormous
subspaces of $\H$ (macroscopic entropy) corresponding to the physical property
that the apparatus pointer is pointing in the $+$ and the $-$ direction,
respectively.  As in all cases where one associates quantum projectors with
macroscopic events, there will be some ambiguity in the precise definition, but
all that matters for the present discussion is that, for all $j$,
\begin{equation} 
	\begin{array}{c} 	P^+|b^+_j\rg = |b^+_j\rg, \quad P^+|b^-_j\rg =
0, \\ 	P^-|b^-_j\rg = |b^-_j\rg, \quad P^-|b^+_j\rg = 0. \\ 	\end{array}
\label{e6.35} 
\end{equation}
Using these definitions, one can work out the weights corresponding to the
families (\ref{e6.14}), (\ref{e6.21}), (\ref{e6.29}), and (\ref{e6.32}).  From
them one obtains the same conditional probabilities as before: (\ref{e6.18}) to
(\ref{e6.20}), (\ref{e6.23}) and (\ref{e6.24}), (\ref{e6.31}), and
(\ref{e6.33}), respectively.  Nor are these probabilities altered if, instead
of assuming complete ignorance, or an initial state $A$ at $t_0$, one
introduces an initial probability distribution which assigns to each $|a_j\rg$
a probability $p_j$ in such a way that the total probability of $A$ is 1.
Thus, while the simplifications employed in (\ref{e6.12}) and the following
discussion make it easier to do the calculations, they do not affect the final
conclusions.

	As a final remark, it may be noted that we have made no use of {\it
decoherence}, in the sense of the interaction of a system with its environment
\cite{zr93}, in discussing measurement problems. This is not to suggest that
decoherence is irrelevant to the theory of quantum measurement; quite the
opposite is the case.  For example, the fact that certain physical properties,
such as pointer positions in a properly designed apparatus, have a certain
stability in the course of time despite perturbations from a random
environment, while other physical properties do not, is a matter of both
theoretical and practical interest.  However, the phenomenon of decoherence
does not, in and of itself, specify which framework is to be employed in
describing a measurement; indeed, in order to understand what decoherence is
all about, one needs to use an appropriate framework.  Hence, decoherence is
not the correct conceptual tool to disentangle conceptual dilemmas brought
about by mixing descriptions from incompatible frameworks.

		\subsection{Three State Paradox}

\label{vaidman}

	Aharonov and Vaidman \cite{av91} (also see Kent
\cite{kt96a}) have introduced a class of paradoxes, of which the following is
the simplest example, in which a particle can be in one of three states:
$|A\rg$, $|B\rg$, or $|C\rg$, and in which the unitary dynamics for a set of
three times $t_0 < t_1 < t_2$ is given by the identity operator: $|A\rg \mt
|A\rg$, etc.  Define
\begin{equation} 
	\begin{array}{c} 	|\Phi\rg = (|A\rg + |B\rg + |C\rg)/\sqrt 3,\\
|\Psi\rg = (|A\rg + |B\rg - |C\rg)/\sqrt 3, 	\end{array}
\label{e6.36} 
\end{equation}
and, consistent with our previous notation, let a letter outside a ket denote
the corresponding projector, and a tilde its complement, thus:
\begin{equation} 
 A=| A\rg\lg A|,\quad \tilde A = I-A = B + C.
\label{e6.37} 
\end{equation}

	Let us begin with the framework based upon
\begin{equation} 
 \IB = \{\Phi_0 + \tilde \Phi_0 \}\{\Psi_2 + \tilde \Psi_2 \},
\label{e6.38} 
\end{equation}
and consider two refinements.  In the first, generated by
\begin{equation} 
 \IB = \{\Phi_0 + \tilde \Phi_0 \}\{A_1 + \tilde A_1\} 	\{\Psi_2 + \tilde
\Psi_2 \},
\label{e6.39} 
\end{equation}  
and easily shown to be consistent, an elementary calculation yields the result:
\begin{equation} 
 \Pr(A_1|\Phi_0\Psi_2) = 1.
\label{e6.40} 
\end{equation}
The second refinement is generated by
\begin{equation} 
 \IB = \{\Phi_0 + \tilde \Phi_0 \}\{B_1 + \tilde B_1\} 	\{\Psi_2 + \tilde
\Psi_2 \},
\label{e6.41}
\end{equation}
and within this framework,
\begin{equation} 
 \Pr(B_1|\Phi_0\Psi_2) = 1.
\label{e6.42} 
\end{equation}

	The paradox comes about by noting that the product of the projectors
$A$ and $B$, and thus $A_1$ and $B_1$, is zero.  Consequently, were $B_1$ an
element of the framework (\ref{e6.39}), (\ref{e6.40}) would imply that
$\Pr(B_1|\Phi_0\Psi_2) = 0$, in direct contradiction to (\ref{e6.42}).  But of
course there is no contradiction when one follows the rules of Sec.~\ref{qr},
because $B_1$ and $A_1$ can never belong to the same refinement of
(\ref{e6.38}).  Thus this paradox is a good illustration of the importance of
paying attention to the framework in order to avoid contradictions when
reasoning about a quantum system, and provides a nice illustration of the
pitfall pointed out at the end of Sec.~\ref{harmony}.


			\section{Some Issues of Interpretation}
\label{interp}

		\subsection{Incompatible Frameworks}
\label{incom}

	The central conceptual difficulty of quantum theory, expressed in the
terminology used in this paper, is the existence of mutually incompatible
frameworks, any one of which can, at least potentially, apply to a particular
physical system, whereas two (or more) cannot be applied to the same system.
Whereas the reasoning procedures described in Sec.~\ref{qr} provide an
internally consistent way of dealing with this ``framework problem'', it is, as
is always the case in quantum theory, very easy to become confused through
habits of mind based upon classical physics.  The material in this section is
intended to address at least some of these sources of confusion at a more
intuitive level, assuming that Sec.~\ref{qr} is sound at the formal level.

	It will be useful to consider the explicit example discussed in
Sec.~\ref{measure}, in which a spin-half particle with $S_x=1/2$ at time $t_0$
is later, at $t_2$, subjected to a measurement of $S_z$, and this measurement
yields the result $S_z=1/2$.  There is then a framework $\Z$, (\ref{e6.29}), in
which one can conclude $Z^+_1$ with probability one: that is, the particle was
in a state $S_z=1/2$ at the intermediate time $t_1$. And there is another,
incompatible, framework $\X$, (\ref{e6.32}), in which, on the basis of the same
initial data, one can conclude $X^+_1$ with probability one: that is, the
particle was in a state $S_x=1/2$ at $t_1$.

	The first issue raised by this example is the following.  The rules of
reasoning in Sec.~\ref{qr} allow us to infer the truth of $Z^+_1$ in framework
$\Z$, and the truth of $X^+_1$ in framework $\X$, but we cannot infer the truth
of $Z^+_1$ {\it and\/} $X^+_1$, because they do not belong to the same
framework.  This is quite different from a classical system, in which we are
accustomed to think that whenever an assertion $E$ is true about a physical
system, in the sense that it can be correctly inferred from some known (or
assumed) data, and $F$ is true in the same sense, then $E$ {\it and\/} $F$ must
be true. As d'Espagnat has emphasized \cite{de87,de89,de95}, this is always a
valid conclusion in standard systems of logic.  But in quantum theory, as
interpreted in this paper, such is no longer the case.  Note that there is no
formal difficulty involved: once we have agreed that quantum mechanics is a
stochastic theory in which the concept of ``true'' corresponds to ``probability
one'', then precisely because probabilities (classical or quantum) only make
sense within some algebra of events, the truth of a quantum proposition is
necessarily labeled, at least implicitly, by that algebra, which in the quantum
case we call a framework.  The existence of incompatible quantum frameworks is
no more or less surprising than the existence of non-commuting operators
representing dynamical variables; indeed, there is a sense in which the former
is a direct consequence of the latter.  Thus physicists who are willing to
accept the basic mathematical framework employed in quantum theory, with its
non-classical non-commutativity, should not be shocked that incompatible
frameworks arise when quantum probabilities are incorporated into the theory in
a consistent, rather than an {\it ad hoc}, manner.  If the dependence of truth
on a framework violates classical intuition, the remedy is to revise that
intuition by working through examples, as in Sec.~\ref{exams}.

	Precisely the same point can be made using the example in
Sec.~\ref{vaidman}.  Indeed, the importance of using the correct framework is
perhaps even clearer in this case, where the projectors $A$ and $B$ commute
with each other.

	A second issue raised by the approach of Sec.~\ref{qr} can be stated in
the form of a question: does quantum theory itself specify a unique framework?
And if the answer is ``no'', as maintained in this paper, does this mean the
interpretation of quantum theory presented here is subjective?  Or that it
somehow implies that physical reality is influenced by the choices made by a
physicist \cite{de89,de95}?

	In response, the first thing to note is that while the choice of
framework is not specified by quantum theory, it is also far from arbitrary.
Thus in our example, given the initial data in the form of $S_x=1/2$ at $t_0$
and the results of the measurement of $S_z$ at $t_2$, $\Z$ is the unique
coarsest framework which contains the data and allows us to discuss the value
of $S_z$ at the time $t_1$.  To be sure, any refinement of this framework would
be equally acceptable, but it is also the case that any refinement would lead
to precisely the same probability of $S_z$ at the time $t_1$, conditional upon
the initial data.  The same holds for the more general situation discussed in
Sec.~\ref{qr}: any refinement of the smallest (coarsest) framework which
contains the data and conclusions will lead to the same probability for the
latter, conditional upon the former.  This is also the case for various sorts
of quantum reasoning constantly employed in practice in order to calculate, for
example, a differential cross section.

	In a certain sense, the very fact that incompatible frameworks are
incompatible is what brings about the quasi-uniqueness in the choice of
frameworks just mentioned.  Certain questions are meaningless unless one uses a
framework in which they mean something, and the same is true of initial data.
Differential scattering cross sections require one type of framework, whereas
the discussion of interference between two parts of a wave going off in
different directions, but later united by a system of mirrors, requires
another.  While this fact is appreciated at an intuitive level by practicing
physicists, they tend to find it confusing, because the general principles of
Sec.~\ref{qr} are not as yet contained in standard textbooks.

	A classical analogy, that of ``coarse graining'' in classical
statistical mechanics, is helpful in seeing why the physicist's freedom in
choosing a quantum framework does not make quantum theory subjective, or imply
that this choice influences physical reality.  As noted in Sec.~\ref{proj},
coarse graining means dividing the classical phase space into a series of cells
of finite volume.  From the point of view of classical mechanics, such a coarse
graining is, of course, arbitrary; cells are chosen because they are convenient
for discussing certain problems, such as macroscopic (thermodynamic)
irreversibility.  But this does not make classical statistical mechanics a
subjective theory.  And, in addition, no one would ever suppose that by
choosing a particular coarse graining, the theoretical physicist is somehow
influencing the system.  If, because it is convenient for his calculations, he
chooses one coarse graining for times $t$ preceding a certain $t_0$, and a
different coarse graining for later times, it would be bizarre to suppose that
this somehow induced a ``change'' in the system at $t_0$!

	To be sure, no classical analogy can adequately represent the quantum
world.  In particular, any two classical coarse grainings are compatible: a
common refinement can always be constructed by using the intersections of cells
from the two families.  And one can always imagine replacing the coarse
grainings by an exact specification of the state of the system.  An analogy
which comes a bit closer to the quantum situation can be constructed by
imposing the rule that one can only use coarse grainings in which the cells
have ``volumes'' which are integer multiples of $h^P$, for a classical system
with $P$ degrees of freedom.  Two coarse grainings which satisfy this condition
will not, in general, have a common refinement which also satisfies this
condition.

	While classical analogies cannot settle things, they are useful in
suggesting ways in which the formalism of Sec.~\ref{qr} can be understood in an
intuitive way.  Eventually, of course, quantum theory, because it is distinctly
different from classical physics, must be understood on its own terms, and an
intuitive understanding of the quantum world must be developed by working
through examples, such as those in Sec.~\ref{exams}, interpreted by means of a
sound and consistent mathematical formalism, such as that of Sec.~\ref{qr}.

	\subsection{Emergence of the Classical World}
\label{qc}

	Both Gell-Mann and Hartle \cite{gmh93}, and Omn\`es \cite{om94b} have
discussed how classical physics expressed in terms of suitable ``hydrodynamic''
variables emerges as an approximation to a fully quantum-mechanical description
of the world when the latter is carried out using suitable frameworks.  While
these two formulations differ somewhat from each other, and from the approach
of the present paper, both are basically compatible with the point of view
found in Secs.~\ref{proj} to \ref{qr}. It is not our purpose to recapitulate or
even summarize the detailed technical discussions by these authors, but instead
to indicate the overall strategy, as viewed from the perspective of this paper,
and comment on how it relates to the problem of incompatible frameworks
discussed above.

	The basic strategy of Gell-Mann and Hartle can be thought of as the
search for a suitable ``quasi-classical'' framework, a consistent family whose
Boolean algebra includes projectors appropriate for representing coarse-grained
variables, such as average density and average momentum inside volume elements
which are not too small, variables which can plausibly be thought of as the
quantum counterparts of properties which enter into hydrodynamic and other
macroscopic descriptions of the world provided by classical physics.  Hence it
is necessary to first find suitable commuting projectors representing
appropriate histories, and then show that the consistency conditions are
satisfied for the corresponding Boolean algebra.  Omn\`es states his strategy
in somewhat different terms which, however, are generally compatible with the
point of view just expressed.

	Both Gell-Mann and Hartle, and Omn\`es, employ consistency conditions
which, unlike those in the present paper, involve a density matrix; see the
discussion in App.~A.  However, the difference is probably of no great
importance when discussing ``quasi-classical'' systems involving large numbers
of particles, for the following reason.  In classical statistical mechanics one
knows (or at least believes!) that for macroscopic systems the choice of
ensemble---microcanonical, canonical, or grand---is for many purposes
unimportant, and, indeed, the average behavior of the ensemble will be quite
close to that of a ``typical'' member. Stated in other words, the use of
probability distributions is a convenience which is not ``in principle''
necessary.  Presumably an analogous result holds for quantum systems of
macroscopic size: the use of a density matrix, both as an ``initial condition''
and as part of the consistency requirement may be convenient, but it is not
absolutely necessary when one is discussing the behavior of a closed system.
For an example in which the final results are to a large degree independent of
what one assumes about the initial conditions, see the discussion at the end of
Sec.~\ref{measure}.

	The task of finding an appropriate quasi-classical consistent family is
made somewhat easier by two facts.  The first is that decoherence \cite{zr93},
in the sense of the interaction of certain degrees of freedom with an
``environment'', can be quite effective in rendering the weight operators
corresponding to minimal elements of a suitably chosen family almost
orthogonal, in the sense discussed in Sec.~\ref{wc}.  (In the present context
one should think of the relevant degrees of freedom as those represented by the
hydrodynamic variables, and the ``environment'' as consisting of the remaining
``microscopic'' variables which are smoothed out, or ignored, in order to
obtain a hydrodynamic description.)  The second is that the weight operators
depend continuously on projectors which form their arguments, and hence it is
at least plausible that if the former are almost orthogonal, small changes in
the projectors can be made in order to achieve exact orthogonality
\cite{dk956}.  Since there is in any case some arbitrariness in choosing
the quantum projectors which represent particular coarse-grained hydrodynamic
variables, small changes in these projectors are unimportant for their physical
interpretation.  Thus exact consistency does not seem difficult to achieve ``in
principle'', even if in practice theoretical physicists are unlikely to be
worried by small deviations from exact orthogonality, as long as these do not
introduce significant inconsistencies into the probabilities calculated from
the weights.  To be sure, there are issues here which deserve further study.

	There are likely to be many different frameworks which are equally good
for the purpose of deriving hydrodynamics from quantum theory, and among these
a number which are mutually incompatible.  Is this a serious problem?  Not
unless one supposes that quantum theory must single out a single framework, a
possibility entertained by Dowker and Kent \cite{dk956}.  If, on the contrary,
the analogy of classical coarse grainings introduced earlier is valid, one
would expect that the same ``coarse-grained'' classical laws would emerge from
any framework which is compatible with this sort of ``quasi-classical''
description of the world.  The internal consistency of the reasoning scheme of
Sec.~\ref{qr}, which can be thought of as always giving the same answer to the
same question, points in this direction, although this is another topic which
deserves additional study.

	There are, of course, many frameworks which are {\it not}
quasi-classical and are incompatible with a ``hydrodynamic'' description of the
world, and there is no principle of quantum theory which excludes the use of
such frameworks.  However, the existence of alternative frameworks does not
invalidate conclusions based upon a quasi-classical framework.  Again, it may
help to think of the analogy of coarse grainings of the classical phase space.
The existence of coarse grainings in which a classical system exhibits no
irreversible behavior---they can be constructed quite easily if one allows the
choice of cells to depend upon the time---does not invalidate conclusions about
thermodynamic irreversibility drawn from a coarse graining chosen to exhibit
this phenomenon.  Similarly, in the quantum case, if we are interested in the
``hydrodynamic'' behavior of the world, we are naturally led to employ
quasi-classical frameworks in which hydrodynamic variables make sense, rather
than alternative frameworks in which such variables are meaningless.

	This suggests an answer to a particular concern raised by Dowker and
Kent \cite{dk956}: If we, as human beings living in a quantum world, have
reason to believe (based upon our memories and the like) that this world has
been ``quasi-classical'' up to now, why should we assume that it will continue
to be so tomorrow?  In order not to be trapped in various philosophical
subtleties such as whether (and if so, how) human thought and belief can be
represented by physical processes, let us consider an easier problem in which
there is a computer inside a closed box, which we as physicists (outside the
box!) have been describing up till now in quasi-classical terms. Suppose,
further, that one of the inputs to the computer is the output of a detector,
also inside the box, measuring radioactive decay of some atoms.  What would
happen if, ten minutes from now, we were to abandon the quasi-classical
framework for one in which, say, there is a coherent quantum superposition of
the computer in distinct macroscopic states?  Of course, nothing particular
would happen to anything inside the box; we, on the other hand, would no longer
be able to describe the object in the box as a computer, because the language
consistent with such a description would be incompatible with the framework we
were using for our description.  The main point can be made using an even
simpler example: consider a spin-half particle in zero magnetic field, and a
history in which $S_x=1/2$ at a time $t_0<t_1$, and $S_z=1/2$ at a time $t_2 >
t_1$.  Nothing at all is happening to the particle at time $t_1$; the only
change is in our manner of describing it.  Additional criticisms of consistent
history ideas with reference to quasi-classical frameworks will be found in
\cite{dk956,kt96b}; responding to them is outside the scope of the present 
paper.


			\section{Conclusion}
\label{summ}

		\subsection{Summary}
\label{sum}

	The counterpart for a closed quantum system of the event space of
classical probability theory is a framework: a Boolean algebra of commuting
projectors on the space $\HB$, (\ref{e2.7}), of quantum histories chosen in
such a way that the weight operators of its minimal elements are orthogonal,
(\ref{e3.14}) or (\ref{e3.14b}).  This ensures that the corresponding weights
are additive, (\ref{e3.11}).  A refinement of a framework is an enlarged
Boolean algebra which again satisfies the consistency conditions.  Two or more
frameworks with a common refinement are called compatible, but in general
different quantum frameworks are incompatible with one another, a situation
which has no classical analog.

	Given some framework and an associated probability distribution, the
rules for quantum reasoning, Sec.~\ref{qr}, are the usual rules for
manipulating probabilities, with ``true'' and ``false'' corresponding to
(conditional) probabilities equal to 1 and 0, respectively.  In addition, a
probability distribution defined on one framework can be extended to a
refinement of this framework using (\ref{e4.5}).  This refinement rule
incorporates the laws of quantum dynamics into the theory: for example, the
Born formula emerges as a conditional probability, (\ref{e5.2}), even in the
absence of any initial data.

	The refinement rule allows descriptions in compatible frameworks to be
combined, or at least compared, in a common refinement.  However, there is no
way of comparing or combining descriptions belonging to incompatible
frameworks, and it is a mistake to think of them as simultaneously applying to
the same physical system.

	Quantum reasoning allows one, on the basis of the same initial data, to
reach different conclusions in different, sometimes mutually incompatible,
refinements.  However, the system is internally consistent in the sense that
the probability assigned to any history on the basis of some initial data
(which must be given in a single framework) is independent of the refinement in
which that history occurs.  Hence it is impossible to conclude that some
consequence of a given set of initial data is both true and false.
Nevertheless, probabilities are only meaningful with reference to particular
frameworks, and the same is the case for ``true'' and ``false'' regarded as
limiting cases in which a probability is 1 or 0.  Hence a basic condition for
sound quantum reasoning is keeping track of the framework employed at a
particular point in an argument.

	\subsection{Open Questions}
\label{open}

	The entire technical discussion in Secs.~\ref{proj} to \ref{qr} is
based upon a finite-dimensional Hilbert space $\H$ for a quantum system at a
single time, and likewise a finite-dimensional history space $\HB$.  This seems
satisfactory for exploring those conceptual difficulties which are already
present in the finite-dimensional case, and allows a simple exposition with a
minimal number of technical conditions and headaches.  And, as a practical
matter, in any situation in which a finite physical system can be thought of as
possessing a finite entropy S, it is reasonable to suppose that the ``right
physics'' will emerge when one restricts one's attention to a subspace of $\H$
with dimension of order $\exp[S/k_B]$.  Nonetheless, introducing such a cutoff,
even for the case of a single particle in a finite box, is mathematically
awkward, and for this reason alone it would be worthwhile to construct the
appropriate extension of the arguments given in this paper to the (or at least
some) infinite-dimensional case.  For some steps in this direction, see the
work of Isham and his collaborators \cite{is94,isl94,isls94}.

	It is not necessary to require that the Boolean algebra of histories
introduced in Sec.~\ref{proj} satisfy the consistency conditions of
Sec.~\ref{wc} in order to introduce a probability distribution on the former.
Consistency becomes an issue only when one considers refinements of a
framework, and wants to define a refined probability.  Even so, one can
introduce refinements of an inconsistent framework $\F$, with probabilities
given by (\ref{e4.5}), by demanding that for each $i$, the weight operator
associated with $F_i$ be a sum of mutually orthogonal weight operators of those
minimal elements $G_j$ of the refinement $\G$ whose sum is $F_i$.  The open
question is whether there is some physical application for such a generalized
system of frameworks and refinement rules.  Consistent frameworks seem to be
sufficient for describing closed quantum systems, but it is possible that
generalized frameworks would be of some use in thinking about an open system: a
subsystem of a closed system in which the remainder of the closed system is
regarded as forming some sort of ``environment'' of the system of interest.

	While the scheme of quantum reasoning presented in this paper has wide
applicability, there are certain to be situations not covered by the rules
given in Sec.~\ref{qr}.  One of these is the case of counterfactuals, such as:
``{\it if} the counter had not been located directly behind the slit, {\it
then} the particle would have $\ldots$''.  Analyzing these requires comparing
two situations which differ in some specific way---e.g., in the position
occupied by some counter---and it is not clear how to embed this in the scheme
discussed in Sec.~\ref{qr}.  Inasmuch as many quantum paradoxes, including some
of the ones associated with double-slit diffraction, and certain derivations of
Bell's inequality and analogous results, make use of counterfactuals, analyzing
them requires considerations which go beyond those in the present paper.  As
philosophers have yet to reach general agreement on a satisfactory scheme for
counterfactual reasoning applied to the classical world
\cite{rhpc}, an extension which covers all of quantum reasoning is likely to be
difficult. On the other hand, one sufficient to handle the special sorts of
counterfactual reasoning found in common quantum paradoxes is perhaps a simpler
problem.

	Can the structure of reasoning developed in this paper for
non-relativistic quantum mechanics be extended to relativistic quantum
mechanics and quantum field theory?  Various examples suggest that the sort of
peculiar non-locality which is often thought to arise from violations of Bell's
inequality and various EPR paradoxes will disappear when one enforces the rules
of reasoning given in Sec.~\ref{qr}. While this is encouraging, it is also the
case that locality (or the lack thereof) in non-relativistic quantum theory has
yet to be carefully analyzed from the perspective presented in this paper, and
hence must be considered among the open questions.  And, of course, getting rid
of spurious non-localities is only a small step along the way towards a fully
relativistic theory.

\section*{Acknowledgements}

	It is a pleasure to acknowledge useful correspondence and/or
conversations with B. d'Espagnat, F. Dowker, S. Goldstein, L. Hardy, J. Hartle,
C. Isham, A.  Kent, R.  Omn\`es, M. Redhead, E. Squires, and L. Vaidman.
Financial support for this research has been provided by the National Science
Foundation through grant PHY-9220726.

\appendix

	\section{Consistency Using a Density Matrix}

\label{appa}

	The consistency condition introduced in Sec.~\ref{wc} differs in a
small but not insignificant way from the one introduced by Gell-Mann and Hartle
\cite{gmh90,gmh93}, based upon a decoherence
functional. The latter employs a density matrix and amounts, in effect, to
replacing the operator inner produce (\ref{e3.8}) by
\begin{equation} 
 \lg A,B\rg = \Tr[A\ad \rho B],
\label{ea.1} 
\end{equation}
where $\rho$ is a density matrix (positive operator with unit trace) or,
\cite{gmh94}, by 
\begin{equation} 
 \lg A,B\rg = \Tr[A\ad \rho B\rho'],
\label{ea.2} 
\end{equation}
where both $\rho$ and $\rho'$ are density matrices, thought of as associated
with the initial and final time, respectively.  Still more general
possibilities have been proposed by Isham et al. \cite{isls94}. While Omn\`es'
approach \cite{om94c} is somewhat different, his consistency condition also
employs a density matrix in a manner similar to (\ref{ea.1}).

	Certainly one cannot object to either (\ref{ea.1}) or (\ref{ea.2}), or
some completely different definition, on purely mathematical grounds.  If, on
the other hand, $\rho$ is to be interpreted as representing something like a
probability distribution for the physical system at an initial time, the
following considerations favor (\ref{e3.8}).  First, given that an arbitrary
probability distribution can be introduced once a framework has been specified,
Sec.~\ref{prob}, and this can refer to the initial time, or the final time, or
to anything in between, there is no (obvious) gain in generality from
introducing a density matrix into the operator inner product.  Second, in the
scheme outlined in Secs.~\ref{proj} to \ref{prob}, the conditions for choosing
a framework are independent of the probability one chooses to assign to the
corresponding histories, whereas employing (\ref{ea.1}) or (\ref{ea.2}) couples
the acceptability of a framework and the probability assigned to its histories
in a somewhat awkward way.  Third, (\ref{e3.8}) is obviously a simpler
construction than either (\ref{ea.1}) or (\ref{ea.2}), and there seems to be no
physical situation in non-relativistic quantum mechanics in which it is not
perfectly adequate.  To be sure, all of these considerations have a certain
aesthetic character, and elegance is not always a good guide to developing a
physical theory, even when there is agreement as to what is most elegant!  The
reader will have to make up his own mind.


\begin{thebibliography}{11}

\bibitem{dv95} D.~P.~DiVincenzo, Science {\bf 270}, 255 (1995).

\bibitem{gn96}
R.~B.~Griffiths and C-S.~Niu, Phys. Rev. Lett. {\bf 76}, 3228 (1996).

\bibitem{bl90} 
J. S. Bell, in {\it Sixty-Two Years of Uncertainty}, edited by A. I. Miller
(Plenum Press, New York, 1990), p. 17.

\bibitem{vn55} 
J. von Neumann {\it Mathematical Foundations of Quantum Mechanics} (Princeton
University Press, Princeton, 1955), Ch. III, Sec. 5.

\bibitem{gr84}
R. B. Griffiths, J. Stat. Phys. {\bf 36}, 219 (1984).

\bibitem{is94}
	C.~J.~Isham, J. Math. Phys. {\bf 35}, 2157 (1994).

\bibitem{gr93}
R. B. Griffiths, Phys. Rev. Lett. {\bf 70}, 2201 (1993).

\bibitem{gr94}
	R. B. Griffiths, in {\it Symposium on the Foundations of Modern Physics
1994}, edited by K.~V.~Laurikainen, C.~Montonen, and K.~Sunnarborg (Editions
Fronti\`eres, Gif-sur-Yvette, France, 1994), p. 85.



\bibitem{om88}
 R. Omn\`es, J. Stat. Phys. {\bf 53}, 893 (1988).

\bibitem{om92}
 R.~Omn\`es, Rev. Mod. Phys. {\bf 64}, 339 (1992).

\bibitem{om94}
R.~Omn\`es, {\it The Interpretation of Quantum Mechanics} (Princeton University
Press, Princeton, 1994).

\bibitem{gmh90}
M. Gell-Mann and J. B. Hartle in {\em Complexity, Entropy, and the Physics of
Information}, edited by W. Zurek (Addison Wesley, Reading, 1990).

\bibitem{gmh93}
M. Gell-Mann and J. B. Hartle, Phys. Rev. D {\bf 47}, 3345 (1993).


\bibitem{gr93b}
R. B. Griffiths, Found. Phys. {\bf 23}, 1601 (1993).

\bibitem{dk956}
F.~Dowker and A.~Kent, Phys. Rev. Lett. {\bf 75}, 3038 (1995); J.~Stat.~Phys.
{\bf 82}, 1575 (1996).

\bibitem{de87} 
B. d'Espagnat, Physics Lett. A {\bf 124}, 204 (1987).

\bibitem{de89} 
B. d'Espagnat, J. Stat. Phys. {\bf 56}, 747 (1989).

\bibitem{de90} 
B. d'Espagnat, Found. Phys. {\bf 20}, 1147 (1990).

\bibitem{de95} 
B. d'Espagnat, {\it Veiled Reality} (Addison-Wesley, Reading, Mass., 1995),
Sec. 11.4.


\bibitem{kt96a}
A.~Kent, ``Consistent Sets Contradict'', preprint: gr-qc/9604012.

\bibitem{av91}
Y.~Aharonov and L.~Vaidman, J. Phys. A {\bf 24}, 2315 (1991); L. Vaidman,
Found. Phys. (to appear), quant-ph/9601005.

\bibitem{fl68}
W.~Feller, {\it An Introduction to Probability Theory and Its Applications},
third edition (John Wiley \& Sons, New York, 1968), Ch.~I.

\bibitem{bvn36} 
G. Birkhoff and J. von Neumann, Annals of Math. {\bf 37}, 823 (1936).

\bibitem{mce95} 
J. N. McElwaine, Phys. Rev. A {\bf 53}, 2021 (1996).

\bibitem{gr84n}
See Sec.~5 of \cite{gr84}

\bibitem{om94b}
Ch.~6 of \cite{om94}

\bibitem{zr93}
W.~H. Zurek, Prog. Theor. Phys. {\bf 89}, 281 (1993). and references given
there.

\bibitem{kt96b}
A.~Kent, to appear in Phys. Rev. A (1996), preprint: gr-qc/9512023

\bibitem{isl94}
C.~J.~Isham and N.~Linden, J.~Math. Phys. {\bf 35}, 5452 (1994).

\bibitem{isls94}
C.~J.~Isham, N.~Linden, and S.~Schreckenberg, J.~Math. Phys. {\bf 35}, 6360
(1994).


\bibitem{rhpc}

	M. Redhead, private communication.

\bibitem{gmh94}
M.~Gell-Mann and J.~B.~Hartle in {\it Physical Origins of Time Asymmetry},
edited by J.~J.~Halliwell, J.~P\'erez-Mercader and W.~H.~Zurek (Cambridge
University Press, Cambridge, 1994), p. 311.

\bibitem{om94c}
Ch.~4 of \cite{om94}.

\end{thebibliography}
\end{document}